\documentclass[reprint,amsmath,amssymb,aps,superscriptaddress,prb]{revtex4-1}
\usepackage{graphicx}
\usepackage{dcolumn}
\usepackage{bm}
\usepackage{epsfig}
\usepackage{soul}
\usepackage{subfigure}
\usepackage{placeins}
\usepackage{chngcntr}
\usepackage[usenames,dvipsnames]{color}

\begin{document}

\preprint{APS/123-QED}

\title{First principles studies of the Gilbert damping and exchange interactions for half-metallic Heuslers alloys}

\date{\today}
\author{Jonathan Chico}
\email{jonathan.chico@physics.uu.se}
\affiliation{Department of Physics and Astronomy, Materials Theory Division, Uppsala University, Box 516, SE-75120 Uppsala, Sweden}

\author{Samara Keshavarz}
\affiliation{Department of Physics and Astronomy, Materials Theory Division, Uppsala University, Box 516, SE-75120 Uppsala, Sweden}

\author{Yaroslav Kvashnin}
\affiliation{Department of Physics and Astronomy, Materials Theory Division, Uppsala University, Box 516, SE-75120 Uppsala, Sweden}

\author{Manuel Pereiro}
\affiliation{Department of Physics and Astronomy, Materials Theory Division, Uppsala University, Box 516, SE-75120 Uppsala, Sweden}

\author{Igor Di Marco}
\affiliation{Department of Physics and Astronomy, Materials Theory Division, Uppsala University, Box 516, SE-75120 Uppsala, Sweden}

\author{Corina Etz}
\affiliation{Department of Engineering Sciences and Mathematics, Materials Science Division, Lule\aa\hspace{1pt}  University of Technology, Lule\aa, Sweden}

\author{Olle Eriksson}
\affiliation{Department of Physics and Astronomy, Materials Theory Division, Uppsala University, Box 516, SE-75120 Uppsala, Sweden}

\author{Anders Bergman}
\affiliation{Department of Physics and Astronomy, Materials Theory Division, Uppsala University, Box 516, SE-75120 Uppsala, Sweden}

\author{Lars Bergqvist}
 \affiliation{Department of Materials and Nano Physics, School of Information and Communication Technology, KTH Royal Institute of Technology, Electrum 229, SE-16440 Kista, Sweden}
 \affiliation{SeRC (Swedish e-Science Research Center), KTH Royal Institute of Technology, SE-10044 Stockholm, Sweden}
 
\begin{abstract}
Heusler alloys have been intensively studied due to the wide variety of properties that they exhibit. One of these properties is of particular interest for technological applications, i.e. the fact that some Heusler alloys are half-metallic. In the following, a systematic study of the magnetic properties of three different Heusler families  $\textrm{Co}_2\textrm{Mn}\textrm{Z}$, $\text{Co}_2\text{Fe}\text{Z}$ and $\textrm{Mn}_2\textrm{V}\textrm{Z}$ with $\text{Z}=\left(\text{Al, Si, Ga, Ge}\right)$ is performed. A key aspect is the determination of the Gilbert damping from first principles calculations, with special focus on the role played by different approximations, the effect that substitutional disorder and temperature effects. Heisenberg exchange interactions and critical temperature for the alloys are also calculated as well as magnon dispersion relations for representative systems, the ferromagnetic $\textrm{Co}_2\textrm{Fe}\textrm{Si}$ and the ferrimagnetic $\textrm{Mn}_2\textrm{V}\textrm{Al}$. Correlations effects beyond standard density-functional theory are treated using both the local spin density approximation including the Hubbard $U$ and the local spin density approximation plus dynamical mean field theory approximation, which allows to determine if dynamical self-energy corrections can remedy some of the inconsistencies which were previously reported for these alloys.
\end{abstract}

\pacs{Valid PACS appear here}
\maketitle


\section{Introduction}

The limitations presented by traditional electronic devices, such as Joule heating, which leads to higher energy consumption, leakage currents and poor scaling with size among others~\cite{awschalom2007challenges}, have sparked profound interest in the fields of spintronics and magnonics. Spintronics applications rely in the transmission of information in both spin and charge degrees of freedom of the electron, whilst in magnonics information is transmitted via magnetic excitations, spin waves or magnons. Half-metallic materials with a large Curie temperature are of great interest for these applications. Due to the fact that they are conductors in only one of the spin channels makes them ideal candidates for possible devices~\cite{RevModPhys.80.315}. Half-metals also have certain advantages for magnonic applications, due to the fact that they are insulators in a spin channel and thus can have a smaller total density of states at the Fermi energy than metals. This can result into a small Gilbert damping, which is an instrumental prerequisite for magnonic applications~\cite{0022-3727-43-26-264001}.

The name \textquotedblleft full Heusler alloys\textquotedblright refer to a set of compounds with formula $\textrm{X}_2 \textrm{Y}\textrm{Z}$ with X and Y typically being transition metals~\cite{Graf20111}. The interest in them stems from the fact that their properties can be completely different from those of their constituents. Heusler compounds can be superconducting~\cite{PhysRevB.31.6971} ($\text{Pd}_2 \text{Y}\text{Sn}$), semiconductors~\cite{0953-8984-13-1-308} (TiCoSb), half-metallic~\cite{PhysRevB.66.174429} ($\text{Co}_2\text{Mn}\text{Si}$), and can show a wide array of magnetic configurations: ferromagnetic~\cite{PhysRevB.66.174429} ($\text{Co}_2\text{Fe}\text{Si}$), ferrimagnetic~\cite{PhysRevB.75.092407} ($\textrm{Mn}_2\text{V}\text{Al}$) or antiferromagnetic~\cite{DEGROOT199145} (CrMnSb). Due to such  a wide variety of behaviours, full Heusler alloys have been studied in great detail since their discovery in 1903, leading to the discovery of new Heusler families such as the half-Heuslers, with formula XYZ, and the inverse Heuslers, with formula  $\text{X}_2 \text{Y}\text{Z}$. The latter tend to exhibit a different crystal structure and have been predicted to show quite remarkable properties~\cite{PhysRevB.87.024420}.

Many Heusler alloys have also been predicted to be half-metallic, in particular $\text{Co}_2\text{Mn}\text{Si}$ has been the focus of many theoretical and experimental works~\cite{0022-3727-43-19-193001,PhysRevB.66.174429,jourdan2014direct}, due to its large Curie temperature of 985 K~\cite{PhysRevB.74.104405}, half-metallicity and low damping parameter, which makes it an ideal candidate for possible spintronic applications. Despite the large amount of research devoted to the half-metallic Heusler alloys, such as Co$_2$MnSi, only recently theoretical predictions of the Gilbert damping parameter have been made for some Heusler alloys~\cite{0022-3727-48-16-164011,Oogane3037}.

In the present work first principle calculations of the full Heusler families $\textrm{Co}_2\textrm{Mn}\textrm{Z}$, $\text{Co}_2\text{Fe}\text{Z}$ and $\textrm{Mn}_2\textrm{V}\textrm{Z}$ with $\text{Z}=\left(\text{Al, Si, Ga, Ge}\right)$ are performed, with special emphasis on the determination of the Gilbert damping and the interatomic exchange interactions. A study treatment of the systems with different exchange correlation potentials is also performed. 

The paper is organized as follows, in section~\ref{sec:AB_INITIO} the computational methods used are presented. Then, in section~\ref{sec:DOS}, magnetic moments and spectral properties are discussed. In section~\ref{sec:EXCHANGE} the results for the exchange stiffness parameter, the critical temperature obtained via Monte Carlo simulations and magnon dispersion relations are presented. Finally in section~\ref{sec:DAMPING}, the calculated damping parameter for the different Heusler is presented and discussed. 

\section{Computational Methods\label{sec:AB_INITIO}}

The full Heusler alloys ($\textrm{X}_2 \textrm{Y}\textrm{Z}$) have a crystal structure given by the space group Fm-3m with $\textrm{X}$ occupying the Wyckoff position 8c ($\frac{1}{4}$,$\frac{1}{4}$,$\frac{1}{4}$), while $\textrm{Y}$ sits in the 4a (0,0,0) and $\textrm{Z}$ in the 4b ($\frac{1}{2}$,$\frac{1}{2}$,$\frac{1}{2}$). 

To determine the properties of the systems first principles electronic structure calculations were performed. They were mainly done by means of the Korringa-Kohn-Rostocker Green's function formalism as implemented in the SPR-KKR package~\cite{SPRKKR}. The shape of the potential was considered by using both the Atomic Sphere Approximation (ASA) and a full potential (FP) scheme. The calculations of exchange interactions were performed in scalar relativistic approximation while the full relativistic Dirac equation was used in the damping calculations. The exchange correlation functional was treated using both the Local Spin Density Approximation (LSDA), as considered by Vosko, Wilk, and Nusair (VWN)~\cite{VWN}, and the Generalized Gradient Approximation (GGA), as devised by Perdew, Burke and Ernzerhof (PBE)~\cite{PhysRevLett.77.3865}. For cases in which substitutional disorder is considered, the Coherent Potential Approximation (CPA) is used~\cite{soven1967pr,stocks1978prl}. 

Static correlation effects beyond LSDA or GGA are taken into account by using the LSDA$+U$ approach, where the Kohn-Sham Hamiltonian is supplemented with an additional term describing local Hubbard interactions~\cite{Hubbard238}, for the $d$-states of Co, Mn and Fe.
The U-matrix describing this on-site interactions was parametrized through the Hubbard parameter $U$ and the Hund exchange $J$, using values $U_\text{Co}=U_\text{Mn}=U_\text{Fe}=3 \text{ eV}$ and $J_\text{Co}=J_\text{Mn}=J_\text{Fe}=0.8 \text{ eV}$, which are in the range of the values considered in previous theoretical studies~\cite{PhysRevB.74.104405,0022-3727-40-6-S12,PhysRevB.87.220402,PhysRevB.72.184434}. This approach is used for the Heusler alloys families $\text{Co}_2\text{Mn}\text{Z}$ and $\text{Co}_2\text{Fe}\text{Z}$, as previous studies have shown that for systems such as $\text{Co}_2\text{Fe}\text{Si}$ it might be necessary to reproduce several experimental observations, although, this topic is still up for debate~\cite{PhysRevB.87.220402}. Since part of correlation effects of the $3d$ orbitals is already included in LSDA, their contribution has to be subtracted before adding the $+U$ self-energy. This contribution to be removed is usually called \textquotedblleft double-counting\textquotedblright (DC) correction and there is no unique way of defining it (see e.g. Ref.~\onlinecite{Karolak-NiO-dc}). We have used two of the most widely used schemes for the DC, namely the Atomic Limit (AL), also known as Fully Localized Limit (FLL)~\cite{PhysRevB.52.R5467}, and the Around Mean Field (AMF) ~\cite{PhysRevB.49.14211}. The dependence of the results on this choice will be extensively discussed in the following sections. 

In order to shine some light on the importance of the dynamical correlations for the magnetic properties of the selected Heusler alloys, a series of calculations were performed in the framework of DFT plus Dynamical Mean Field Theory (DMFT)\cite{dmft-orig,dft-dmft}, as implemented in the full-potential linear muffin-tin orbital (FP-LMTO) code RSPt~\cite{rspt}. 
As for LSDA$+U$, the DMFT calculations are performed for a selected set of metal $3d$ orbitals on top of the LSDA solution in a fully charge self-consistent manner.\cite{csc-dmft,grechnev-FeCoNi-PRB}
The effective impurity problem, which is the core of the DMFT, is solved through the spin-polarized T-matrix fluctuation-exchange (SPTF) solver~\cite{sptf}. 
This type of solver is perturbative and is appropriate for the systems with moderate correlation effects, where $U/W<1$ ($W$ denotes the bandwidth).\cite{ldapp}
Contrary to the prior DMFT studies\cite{chioncel-cms,chadov-dmft}, we have performed the perturbation expansion of the Hartree-Fock-renormalized Green's function ($G_{HF}$) and not of the bare one. 
Concerning the DC correction, we here use both the FLL approach, described above, as well as the so-called \textquotedblleft$\Sigma(0)$\textquotedblright correction. In the latter case, the orbitally-averaged static part of the DMFT self-energy is removed, which is often a good choice for metals\cite{DMFT-Fe-mag,dft-dmft}. Finally, in order to extract information about the magnetic excitations in the systems, we have performed a mapping onto an effective Heisenberg Hamiltonian
\begin{equation}
\hat H = -\sum_{i \ne j} J_{ij} \vec e_i \vec e_j ,
\label{H-M}
\end{equation}
where $J_{ij}$ is an exchange interaction between the spins located at site \textit{i} and \textit{j}, while the $\vec e_i$ ($\vec e_j$) represents the unity vector along the magnetization direction at site \textit{i (j)}. The exchange parameters then are computed by making use of the well established LKAG (Liechtenstein, Katsnelson, Antropov, and Gubanov) formalism, which is based on the magnetic force theorem~\cite{Liechtenstein198765,0305-4608-14-7-007,liechten2}. More specific details about the implementation of the LKAG formalism in RSPt can be found in Ref.~\onlinecite{jij-yar}. We also note that the performance of the RSPt method was recently published in Ref.~\cite{Lejaeghereaad3000} and it was found that the accuracy was similar to that of augmented plane wave methods.

From the exchange interactions between magnetic atoms, it is possible to obtain the spin wave stiffness, $D$, which, for cubic systems is written as~\cite{pajda2001prb}
\begin{equation}
  D=\frac{2}{3}\sum_{i,j}\frac{J_{ij}}{\sqrt{m_i m_j}}\left|r_{ij}\right|^2\exp\left(-\eta\frac{r_{ij}}{a_\textrm{lat}}\right),
  \label{eq:stiff}
\end{equation}
where the $m_i$'s are the magnetic moments of a given atom, $r_{ij}$ is the distance between the two considered magnetic moments, $a_\text{lat}$ is the lattice parameter, $\eta$ is a convergence parameter used to ensure the convergence of  Eq.~\ref{eq:stiff}, the value of $D$ is taken under the limit $\eta\rightarrow0$. To ensure the convergence of the summation, it is also important to take into consideration long range interactions. Hence the exchange interactions are considered up to 6 lattice constants from the central atom.

The obtained exchange interactions were then used to calculate the critical temperature by making use of the Binder cumulant, obtained from Monte Carlo simulations as implemented in the UppASD package~\cite{skubic2008jpcm}. This was calculated for three different number of cell repetitions (10x10x10, 15x15x15 and 20x20x20), with the intersection point determining the critical temperature of the system~\cite{0034-4885-60-5-001}.

The Gilbert damping, $\alpha$, is calculated via linear response theory~\cite{PhysRevLett.107.066603}. Temperature effects in the scattering process of electrons are taken into account by considering an alloy analogy model within CPA with respect to the atomic displacements and thermal fluctuations of the spin moments~\cite{ebert2015prb}. Vertex corrections are also considered here, because they provide the \textquotedblleft scattering in\textquotedblright  term of the Boltzmann equation and it corrects significant error in the damping, whenever there is an appreciable s-p or s-d scattering in the system~\cite{Butler1985,SPRKKR}.

From the calculated exchange interactions, the adiabatic magnon spectra (AMS) can be determined by calculating the Fourier transform of the interatomic exchange interactions~\cite{PhysRevB.58.293}. This is determined for selected cases and is compared with the magnon dispersion relation obtained from the dynamical structure factor, $S^k\left(\mathbf{q},\omega\right)$, resulting from spin dynamics calculations. The $S^k\left(\mathbf{q},\omega\right)$ is obtained from the Fourier transform of the time and spatially displaced spin-spin correlation function, $C^k\left(\mathbf{r}-\mathbf{r}',t\right)$~\cite{0953-8984-27-24-243202}

\begin{equation}
\displaystyle
S^k\left(\mathbf{q},\omega\right)=\frac{1}{\sqrt{2\pi}N}\sum_{\mathbf{r},\mathbf{r}'} e^{i\mathbf{q}\cdot\left(\mathbf{r}-\mathbf{r}'\right)}\int_{-\infty}^\infty e^{i\omega t} C^k\left(\mathbf{r}-\mathbf{r}',t\right)\text{d}t.
\end{equation}

The advantage of using the dynamical structure factor over the adiabatic magnon spectra is the capability of studying temperature effects as well as the influence of the damping parameter determined from first principles calculations or from experimental measurements.

\section{Electronic structure\label{sec:DOS}}

The calculated spin magnetic moments for the selected systems are reported in Table~\ref{tab:mom_FM}. These values are  obtained from SPR-KKR with various approximations of the exchange correlation potential and for different geometrical shapes of the potential itself. For the $\textrm{Co}_2\textrm{Mn}\textrm{Z}$ family, when $\text{Z}=\left(\text{Si}, \text{ Ge}\right)$, the obtained spin magnetic moments do not seem to be heavily influenced by the choice of exchange correlation potential or potential shape. However, for $\text{Z}=\left(\text{Al}, \text{ Ga}\right)$ a large variation is observed in the spin moment when one includes the Hubbard parameter $U$. 

\begin{table*}[]
    \centering
    \caption{Summary of the spin magnetic moments obtained using different approximations as obtained from SPR-KKR for the $\textrm{Co}_2\textrm{Mn}\textrm{Z}$ and $\textrm{Co}_2\textrm{Fe}\textrm{Z}$ families with $ \textrm{Z}=\left(\text{Al},\text{Si},\text{Ga},\text{Ge}\right)$. Different exchange correlation potential approximations and shapes of the potential have been used. The symbol $^\dagger$ signifies that the Fermi energy is located at a gap in one of the spin channels.\label{tab:mom_FM}}
    \begin{tabular}{l c c c c c c c c}
    \toprule
    \textbf{Quantity}					&$\text{Co}_2\text{Mn}\text{Al}$	& $\text{Co}_2\text{Mn}\text{Ga}$	& $ \text{Co}_2 \text{Mn}\text{Si}$ 	& $\text{Co}_2\text{Mn}\text{Ge}$   &$\text{Co}_2\text{Fe}\text{Al}$& $\text{Co}_2\text{Fe}\text{Ga}$	& $ \text{Co}_2 \text{Fe}\text{Si}$ 	& $\text{Co}_2\text{Fe}\text{Ge}$\\ \hline
    $a_\text{lat}$ [\AA]					& 5.75~\cite{Buschow198190}		& 5.77~\cite{Buschow198190}	  	    & 5.65~\cite{Webster19711221}		& 5.743~\cite{JAP875463}            & 5.730~\cite{Buschow198190}        &5.737~\cite{Buschow198190}  	        & 5.640~\cite{PhysRevB.87.220402}	& 5.750~\cite{0022-3727-40-6-S01}\\ 
    $m_\text{LDA}^{ASA}$ $[\mu_B]$			& 4.04$^\dagger$ 	        & 4.09$^\dagger$		            & 4.99$^\dagger$ 	                & 4.94$^\dagger$                              			& 4.86$^\dagger$					& 4.93$^\dagger$		        & 5.09 				            & 5.29 				            \\ 
    $m_\text{GGA}^{ASA}$ $[\mu_B]$			& 4.09$^\dagger$		        & 4.15$^\dagger$ 	            & 4.99$^\dagger$	    	                & 4.96$^\dagger$                              			& 4.93$^\dagger$					& 5.00$^\dagger$		        & 5.37							& 5.53            			    \\ 
    $m_{\text{LDA}+U}^{ASA}$ AMF $[\mu_B]$	& 4.02$^\dagger$				& 4.08							& 4.98$^\dagger$						& 4.98$^\dagger$                              			& 4.94$^\dagger$					& 4.99$^\dagger$				& 5.19							& 5.30							\\ 
    $m_{\text{LDA}+U}^{ASA}$ FLL $[\mu_B]$	& 4.77						& 4.90							& 5.02$^\dagger$						& 5.11                              			& 5.22							& 5.36 				        & 5.86$^\dagger$					& 5.94$^\dagger$					\\ 
    $m_\text{LDA}^{FP}$ $[\mu_B]$			& 4.02$^\dagger$				& 4.08$^\dagger$					& 4.98$^\dagger$						& 4.98$^\dagger$                                			& 4.91$^\dagger$					& 4.97$^\dagger$				& 5.28							& 5.42							\\ 
    $m_\text{GGA}^{FP}$ $[\mu_B]$			& 4.03$^\dagger$				& 4.11							& 4.98$^\dagger$						& 4.99$^\dagger$                                			& 4.98$^\dagger$					& 5.01$^\dagger$				& 5.55							& 5.70							\\       
    $m_{\text{LDA}+U}^{FP}$ AMF $[\mu_B]$	& 4.59 						& 4.99							& 4.98$^\dagger$						& 5.13                              			& 5.12							& 5.40						& 5.98$^\dagger$					& 5.98$^\dagger$					\\ 
    $m_{\text{LDA}+U}^{FP}$ FLL $[\mu_B]$	& 4.03$^\dagger$				& 4.17							& 4.99$^\dagger$						& 4.99$^\dagger$                              			& 4.99$^\dagger$					& 5.09						& 5.86$^\dagger$					& 5.98$^\dagger$					\\ 
    $m_\text{exp}$ $[\mu_B]$				& 4.04~\cite{Buschow19831}		& 4.09~\cite{PhysRevB.74.172412}	& 4.96~\cite{0953-8984-12-8-325}		& 4.84~\cite{0953-8984-12-8-325}			& 4.96~\cite{Buschow19831}		& 5.15~\cite{0953-8984-12-8-325}	& 6.00~\cite{PhysRevB.72.184434}		& 5.74~\cite{5257105}            				\\ \toprule
    \end{tabular}
\end{table*}

For the $\textrm{Co}_2\textrm{Fe}\textrm{Z}$ systems, a pronounced difference can be observed in the magnetic moments between the LSDA and the experimental values for $\text{Z}=\left(\text{Si}, \text{ Ge}\right)$. Previous theoretical works~\cite{PhysRevB.74.104405,0022-3727-40-6-S12,PhysRevB.72.184434} suggested that the inclusion of a $+U$ term is necessary to obtain the expected spin magnetic moments, but such a conclusion has been recently questioned~\cite{PhysRevB.87.220402}. To estimate which double counting scheme would be most suitable to treat correlation effects in this class of systems, an interpolation scheme between the FLL and AMF treatments was tested, as described in Ref.~\onlinecite{PhysRevB.67.153106} and implemented in the FP-LAPW package Elk~\cite{ELK}. It was found that both Co$_2$MnSi and Co$_2$FeSi are better described with the AMF scheme, as indicated by their small $\alpha_U$ parameter of $\sim 0.1$ for both materials ($\alpha_U =0$ denotes complete AMF and $\alpha_U=1$ FLL), which is in agreement with the recent work by Tsirogiannis and Galanakis~\cite{Tsirogiannis2015297}.

To test whether a more sophisticated way to treat correlation effects improves the description of these materials, electronic structure calculations for Co$_2$MnSi and Co$_2$FeSi using the DMFT scheme were performed. 
The LSDA+DMFT[$\Sigma(0)$] calculations yielded total spin moments of 5.00~$\mu_B$ and 5.34~$\mu_B$ for respectively Co$_2$MnSi and Co$_2$FeSi. 
These values are almost equal to those obtained in LSDA, which is also the case in elemental transition metals~\cite{grechnev-FeCoNi-PRB}. As mentioned above for LSDA$+U$, the choice of the DC is crucial for these systems. The main reason why no significant differences are found between DMFT and LSDA values is that the employed \textquotedblleft$\Sigma(0)$\textquotedblright DC almost entirely preserves the static part of the exchange splitting obtained in LSDA\footnote{If one utilizes the \textquotedblleft $\Sigma(0)$\textquotedblright DC for SPTF performed for the bare Green's function, then the static terms are strictly preserved. In our case there is some renormalization due to the Hartree-Fock renormalization.}.
For instance, by using FLL DC, we obtained a total magnetization of 5.00~$\mu_B$ and 5.61~$\mu_B$ in Co$_2$MnSi and Co$_2$FeSi, respectively.
We note that the spin moment of Co$_2$FeSi still does not reach the value expected from the Slater-Pauling rule, but the DMFT modifies it in a right direction, if albeit to a smaller degree that the LSDA$+U$ schemes.

\begin{figure*}
    \centering
    \includegraphics[width=0.45\textwidth]{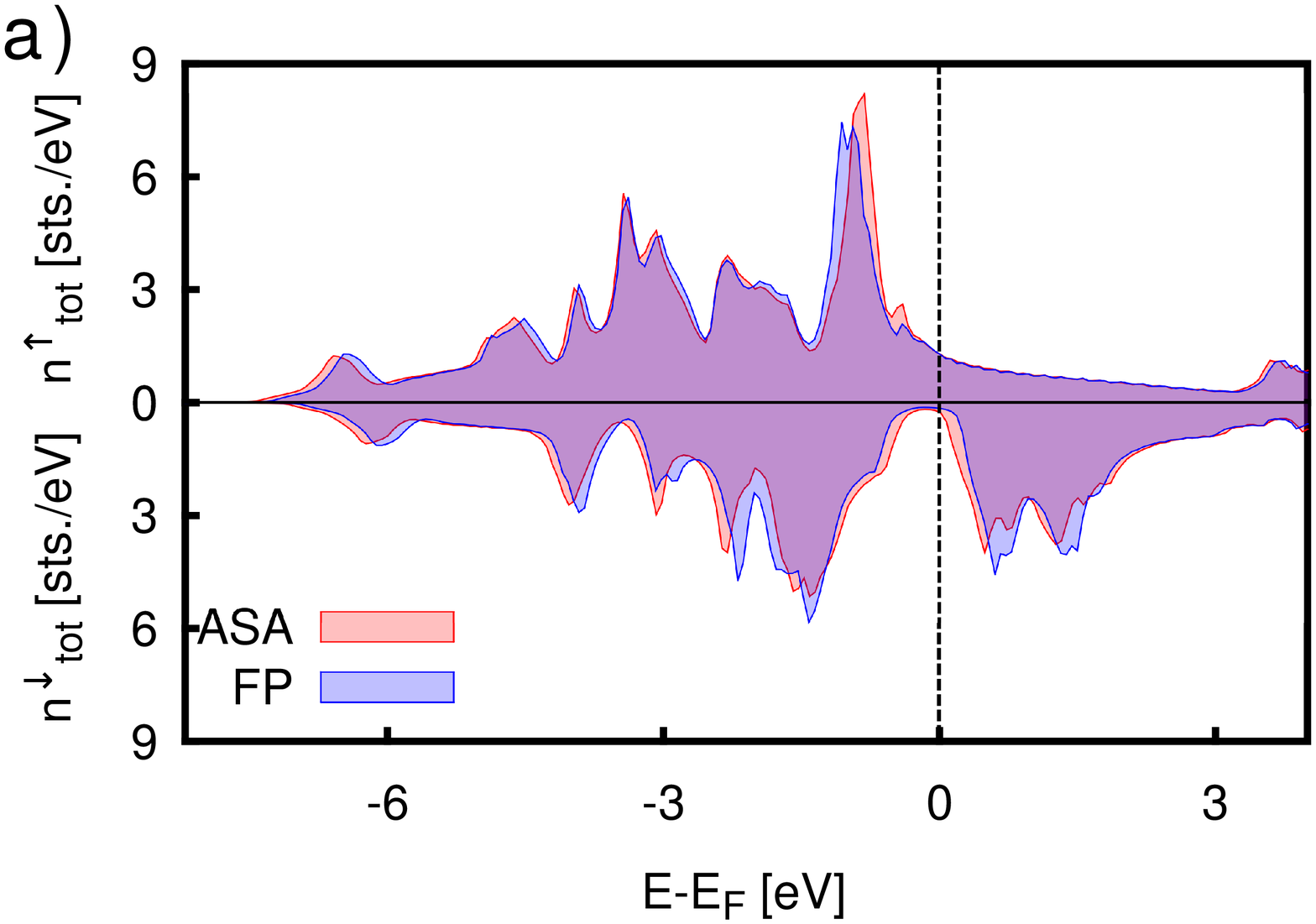}
    \includegraphics[width=0.45\textwidth]{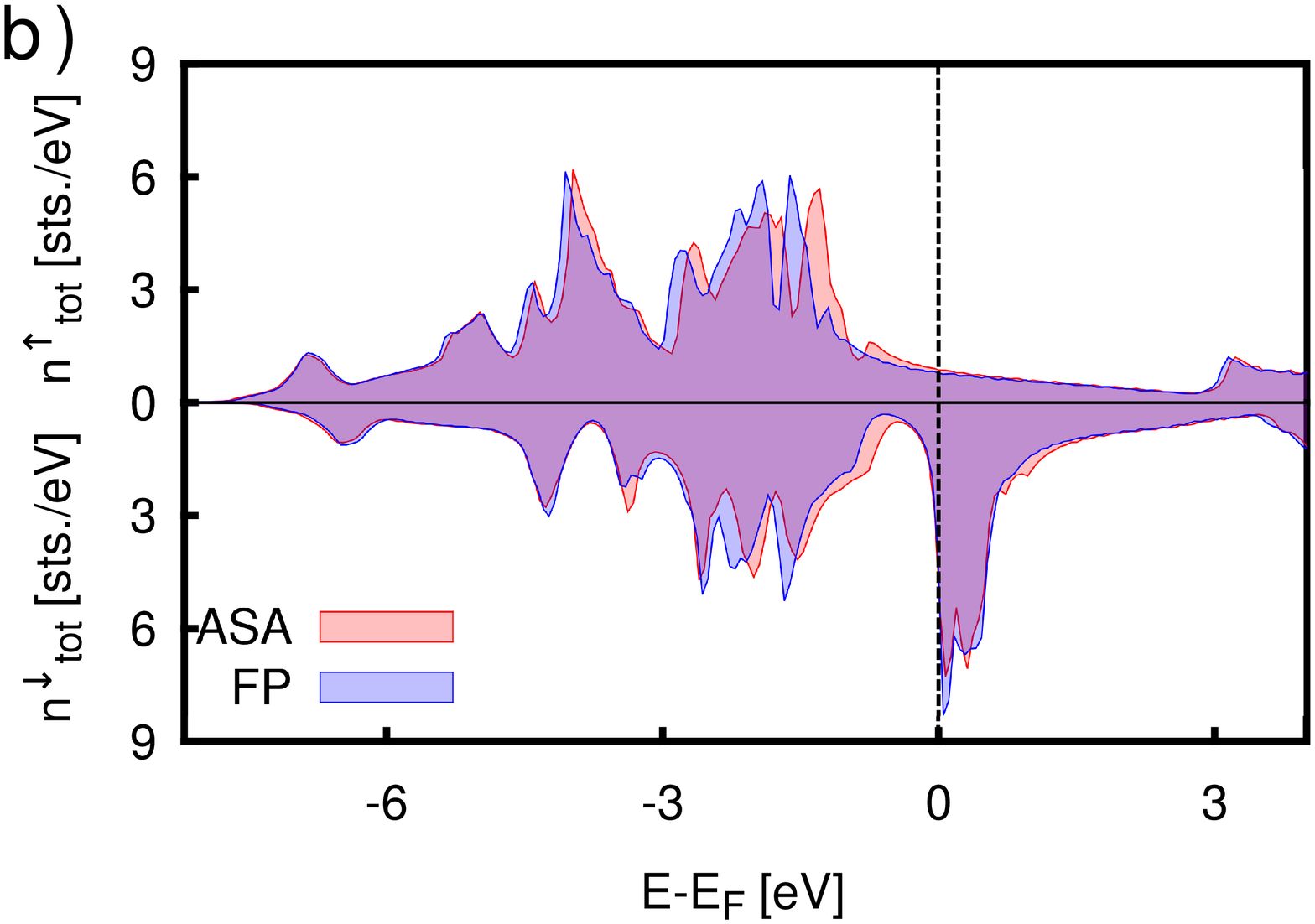}\\
    \includegraphics[width=0.45\textwidth]{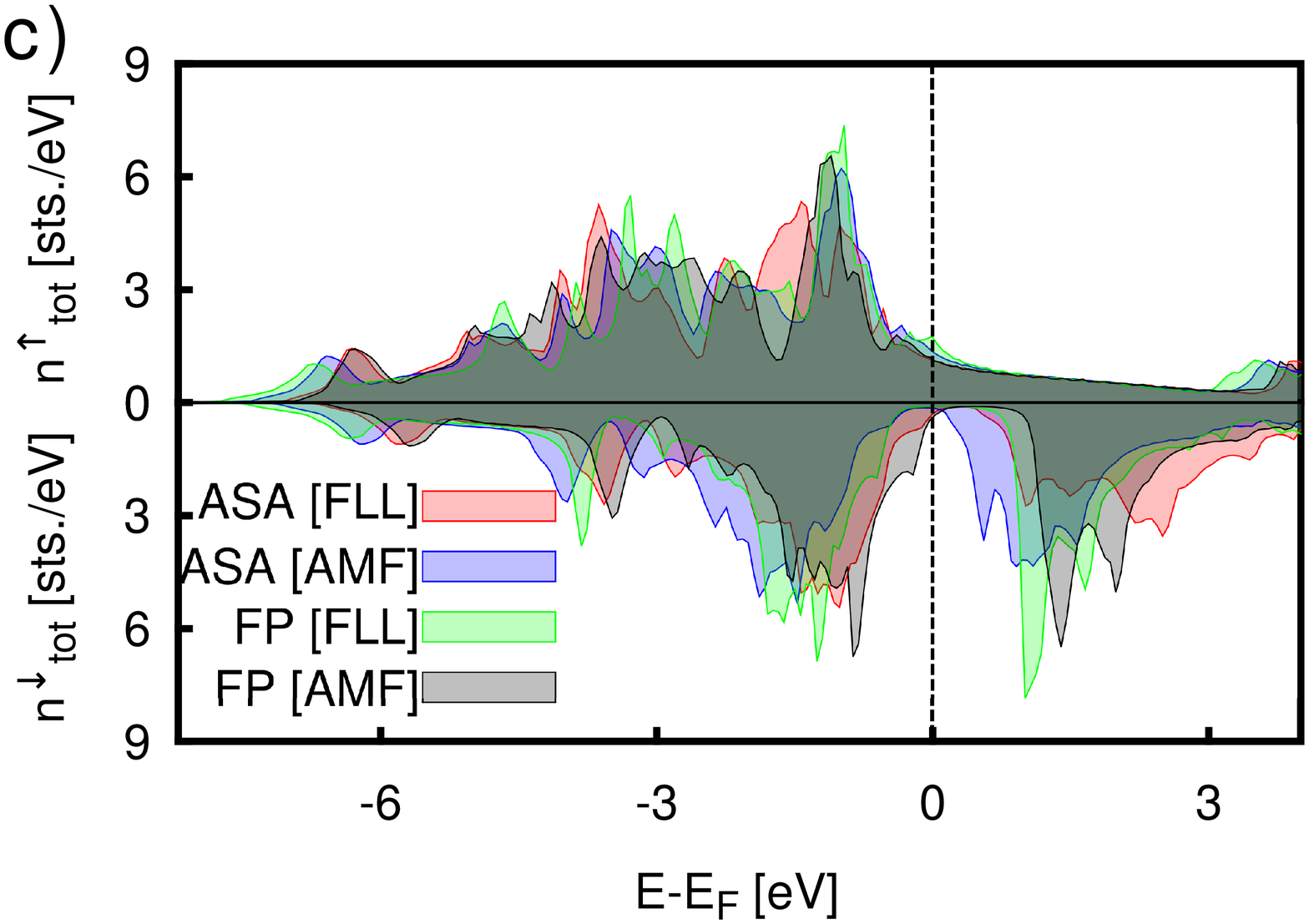}
    \includegraphics[width=0.45\textwidth]{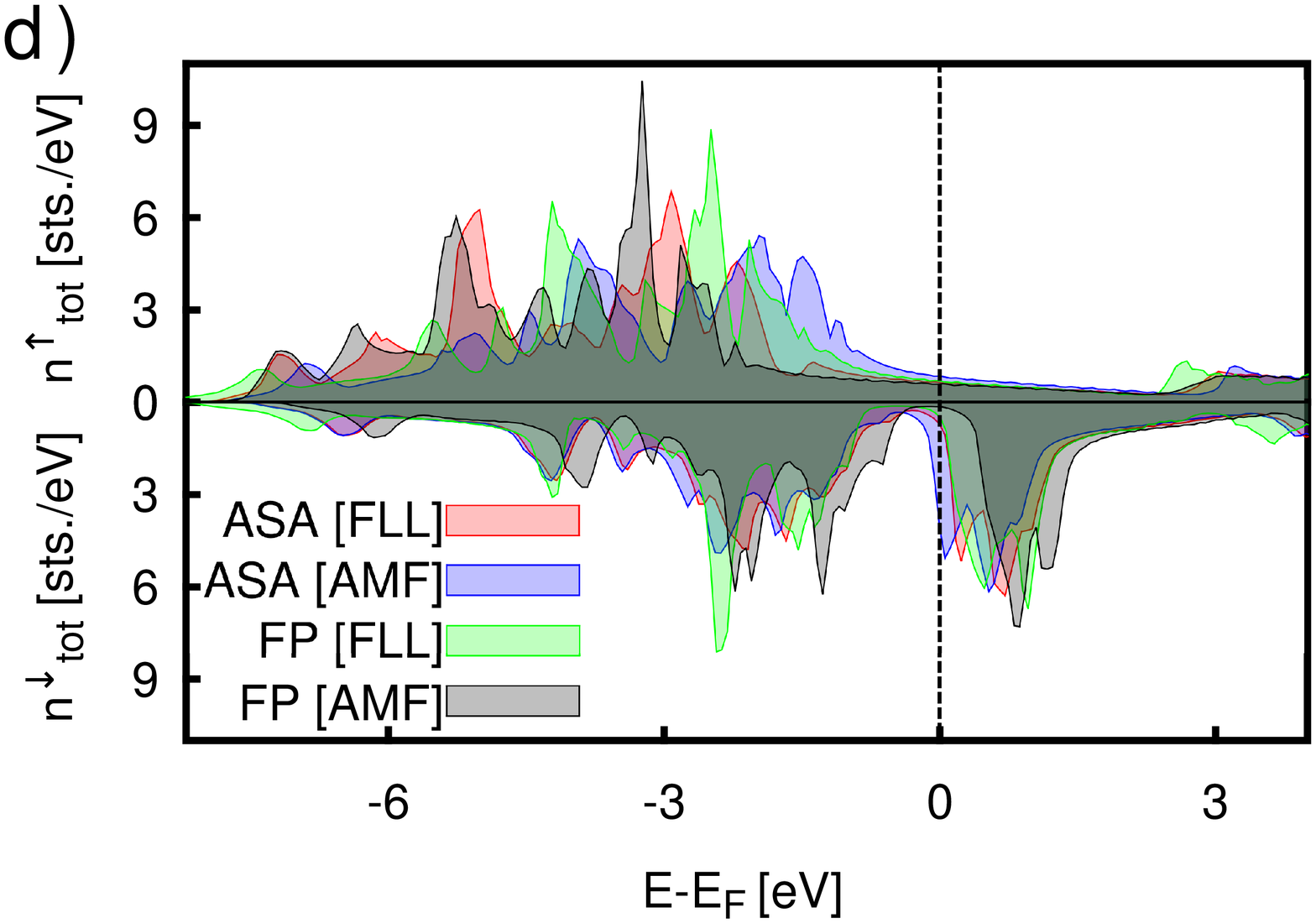}
    \caption{(Color online) Total density of states for different exchange correlation potentials with the dashed line indicating the Fermi energy, sub-figures a) and b) when LSDA is used for $\text{Co}_2\text{Mn}\text{Si}$ and $\text{Co}_2\text{Fe}\text{Si}$ respectively. Sub-figures c) and d) show the DOS when the systems (Co$_2$MnSi and Co$_2$FeSi respectively) are treated with LSDA$+U$. It can be seen that the half metalicity of the materials can be affected by the shape of the potential and the choice of exchange correlation potential chosen.\label{fig:DOS_CFS} }
\end{figure*}

Another important aspect of the presently studied systems is the fact that they are predicted to be half-metallic. In Fig.~\ref{fig:DOS_CFS}, the density of states (DOS) for both Co$_2$MnSi and Co$_2$FeSi is presented using LSDA and LSDA$+U$. For Co$_2$MnSi, the DOS at the Fermi energy is observed to exhibit a very clear gap in one of the spin channels, in agreement with previous theoretical works~\cite{PhysRevB.66.174429}. For Co$_2$FeSi, instead a small pseudo-gap region is observed in one of the spin channels, but the Fermi level is located just at the edge of the boundary as shown in previous works~\cite{PhysRevB.72.184434}. Panels a) and b) of Fig.~\ref{fig:DOS_CFS} also show that some small differences arise depending on the ASA or FP treatment. In particular, the gap in the minority spin channel is slightly reduced in ASA.

When correlation effects are considered within the LSDA$+U$ method, the observed band gap for Co$_2$MnSi becomes larger, while the Fermi level is shifted and still remains in the gap. When applying LSDA$+U$ to Co$_2$FeSi in the FLL scheme, $E_F$  is shifted farther away from the edge of the gap, which explains why the moment becomes almost an integer as expected from the Slater-Pauling behaviour~\cite{PhysRevB.66.174429,PhysRevB.72.184415,PhysRevB.72.184434}. Moreover, one can see that in ASA the gap in the spin down channel is much smaller in comparison to the results obtained in FP.

When the dynamical correlation effects are considered via DMFT, the overall shape of DOS remains to be quite similar to that of bare LSDA, especially close to the Fermi level, as seen in Fig.~\ref{fig:DOS_APP} in the Appendix~\ref{app:DOSDMFT}. 
This is related to the fact that we use a perturbative treatment of the many-body effects, which favours Fermi-liquid behaviour.
Similarly to LSDA+$U$, the LSDA+DMFT calculations result in the increased spin-down gaps, but the produced shift of the bands is not as large as in LSDA+$U$.
This is quite natural, since the inclusion of the dynamical correlations usually tends to screen the static contributions coming from LSDA+$U$.

According to Ref.~\onlinecite{chioncel-cms} taking into account dynamical correlations in Co$_2$MnSi results in the emergence of the non-quasiparticle states (NQS's) inside the minority-spin gap, which at finite temperature tend to decrease the spin polarisation at the Fermi level. 
These NQS's were first predicted theoretically for model systems\cite{irkhin-nqs} and stem from the electron-magnon interactions, which are accounted in DMFT (for review, see Ref.~\onlinecite{RevModPhys.80.315}).
Our LSDA+DMFT results for Co$_2$MnSi indeed show the appearance of the NQS's, as evident from the pronounced imaginary part of the self-energy at the bottom of the conduction minority-spin band (see Appendix~\ref{app_NQS}).
An analysis of the orbital decomposition of the self-energy reveals that the largest contribution to the NQS's comes from the Mn-$T_{Eg}$ states.
However, in our calculations, where the temperature was set to 300K, the NQS's appeared above Fermi level and did not contribute to the system's depolarization, in agreement with the recent experimental study\cite{jourdan2014direct}.

We note that a half-metallic state with a magnetic moment of around 6 $\mu_B$ for Co$_2$FeSi was reported in a previous LSDA+DMFT[FLL] study by Chadov~\textit{et al.\cite{chadov-dmft}}.
In their calculations, both LSDA+$U$ and LSDA+DMFT calculations resulted in practically the same positions of the unoccupied spin-down bands, shifted to the higher energies as compared to LSDA. 
This is due to technical differences in the treatment of the Hartree-Fock contributions to the SPTF self-energy, which in Ref.~\onlinecite{chadov-dmft} is done separately from the dynamical contributions, while in this study a unified approach is used. Overall, the improvements in computational accuracy with respect to previous implementations could be responsible for the obtained qualitative disagreement with respect to Refs.~\onlinecite{chadov-dmft,chioncel-cms}.
Moreover, given that the results qualitatively depend on the choice of the DC term, the description of the electronic structure of Co$_2$FeSi is not conclusive.

The discrepancies in the magnetic moments presented in Table~\ref{tab:mom_FM} with respect to the experimental values can in part be traced back to details of the density of the states around the Fermi energy. The studied Heusler alloys are thought to be half-metallic, which in turn lead to integer moments following the Slater-Pauling rule~\cite{PhysRevB.66.174429}. Therefore, any approximation that destroys half-metallicity will have a profound effect on their magnetic properties~\cite{PhysRevB.66.174429}. For example, for Co$_2$FeAl when the potential is treated in LSDA$+U$[FLL] with ASA the Fermi energy is located at a sharp peak close to the edge of the band gap, destroying the half-metallic state (See supplementary material Fig.1). A similar situation occurs in LSDA$+U$[AMF] with a full potential scheme. It is also worth mentioning that despite the fact that the Fermi energy for many of these alloys is located inside the pseudo-gap in one of the spin channels, this does not ensure a full spin polarization, which is instead observed in systems as e.g. Co$_2$MnSi. Another important factor is the fact that $E_F$ can be close to the edge of the gap as in Co$_2$MnGa when the shape of the potential is considered to be given by ASA and the exchange correlation potential is dictated by LSDA, hence the half-metallicity of these alloys could be destroyed due to temperature effects.

The other Heusler family investigated here is the ferrimagnetic $\text{Mn}_2\text{V}\text{Z}$ with $\textrm{Z}=\left(\text{Al},\text{Si},\text{Ga},\text{Ge}\right)$. The lattice constants used in the simulations correspond to either experimental or previous theoretical works. These data are reported in Table~\ref{tab:mom_Ferri} together with appropriate references. Table~\ref{tab:mom_Ferri} also illustrates the magnetic moments calculated using different exchange correlation potentials and shapes of the potential. It can be seen that in general there is a good agreement with previous works, resulting in spin moments which obey the Slater-Pauling behaviour.

\begin{table}[]
\caption{Lattice constants used for the electronic structure calculations and summary of the magnetic properties for $\textrm{Mn}_2\textrm{V}\textrm{Z}$ with $\textrm{Z}=\left( \text{Al},\text{Si},\text{Ga},\text{Ge}\right)$. As for the ferromagnetic families, different shapes of the potential and exchange correlations potential functionals were used. The magnetic moments follow quite well the Slater-Pauling behavior with all the studied exchange correlation potentials. The symbol $^\dagger$ signifies that the Fermi energy is located at a gap in one of the spin channels.\label{tab:mom_Ferri}}
\begin{tabular}{l c c c c }
 \toprule
 \textbf{Quantity}                          &$\text{Mn}_2\text{V}\text{Al}$	& $\text{Mn}_2\text{V}\text{Ga}$		& $ \text{Mn}_2 \text{V}\text{Si}$ 	& $\text{Mn}_2\text{V}\text{Ge}$ 	\\ \hline
 $a_\text{lat}$ [\AA]		                & 5.687~\cite{Ozdogan2006}		& 5.905~\cite{RameshKumar20082737}	& 6.06~\cite{Ozdogan2006}	&		6.095~\cite{0953-8984-17-6-017}   \\ 
 $m_\text{LDA}^{ASA}$ $[\mu_B]$           	& 1.87  		                		& 1.97$^\dagger$	                    	& 1.00$^\dagger$	                    & 0.99$^\dagger$ 	\\
 $m_\text{GGA}^{ASA}$ $[\mu_B]$	            & 1.99$^\dagger$		            & 2.04$^\dagger$	                    	& 1.01$^\dagger$		                & 1.00$^\dagger$ 	\\ 
 $m_\text{LDA}^{FP}$ $[\mu_B]$           	& 1.92			                & 1.95$^\dagger$	                    	& 0.99$^\dagger$	                    & 0.99 	\\ 
 $m_\text{GGA}^{FP}$ $[\mu_B]$           	& 1.98$^\dagger$	                & 2.02$^\dagger$	                    	& 0.99$^\dagger$	                    & 0.99$^\dagger$	\\ 
 $m_\text{exp}$ $[\mu_B]$               		& ---				            & 1.86~\cite{RameshKumar20082737}	& ---				                & ---	    \\ \toprule
\end{tabular}
\end{table}

For these systems, the Mn atoms align themselves in an anti-parallel orientation with respect to the V moments, resulting in a ferrimagnetic ground state. As for the ferromagnetic compounds, the DOS shows a pseudogap in one of the spin channels (see supplementary material Fig.8-9) indicating that at $T=0 \text{ K}$ these compounds could be half-metallic. An important factor is the fact that the spin polarization for these systems is usually considered to be in the opposite spin channel than for the ferromagnetic alloys presently studied, henceforth the total magnetic moment is usually assigned to a negative sign such that it complies with the Slater-Pauling rule~\cite{PhysRevB.66.174429,Ozdogan2006}.

\section{Exchange interactions and magnons\label{sec:EXCHANGE}}

In this section, the effects that different exchange correlation potentials and geometrical shapes of the potential have over the exchange interactions will be discussed. 

\subsection{Ferromagnetic $\textrm{Co}_2\textrm{Mn}\textrm{Z}$ and $\textrm{Co}_2\textrm{Fe}\textrm{Z}$ with $  \textrm{Z}=\left(\text{Al},\text{Si},\text{Ga},\text{Ge}\right)$}

In Table~\ref{tab:jxc_HeuslerFM} the calculated spin wave stiffness, $D$, is shown. In general there is a good agreement between the calculated values for the $\text{Co}_2\text{Mn}\text{Z}$ family, with the obtained values using LSDA or GGA being somewhat larger than the experimental measurements. This is in agreement with the observations in the previous section, in which the same exchange correlation potentials were found to be able to reproduce the magnetic moments and half-metallic behaviour for the $\text{Co}_2\text{Mn}\text{Z}$ family. In particular, for Co$_2$MnSi the ASA calculations are in agreement with experiments~\cite{0022-3727-42-8-084005,JAP111023912} and previous theoretical calculations~\cite{thoene2009jpd}. It is important to notice that the experimental measurements are performed at room temperature, which can lead to softening of the magnon spectra, leading to a reduced spin wave stiffness.

However, for the $\text{Co}_2\text{Fe}\text{Z}$ family neither LSDA or GGA can consistently predict the spin wave stiffness, with Z=(Al, Ga) resulting in an overestimated value of $D$, while for Co$_2$FeSi the obtained value is severely underestimated. However, for some materials in this family, e.g. Co$_2$FeGa the spin wave stiffness agrees with previous theoretical results~\cite{thoene2009jpd}. These data reflect the influence that certain approximations have on the location of the Fermi level, which previously has been shown to have profound effects on the magnitude of the exchange interactions~\cite{0953-8984-19-43-436227}. This can be observed in the half-metallic Co$_2$MnSi; when it is treated with LSDA$+U$[FLL] in ASA the Fermi level is located at the edge of the gap (see Fig.~\ref{fig:DOS_CFS}c). Resulting in a severely underestimated spin wave stiffness with respect to both the LSDA value and the experimental measurements (see Table~\ref{tab:jxc_HeuslerFM}). The great importance of the location of the Fermi energy on the magnetic properties can be seen in the cases of Co$_2$MnAl and Co$_2$MnGa. In LSDA$+U$[FLL], these systems show non integer moments which are overestimated with respect to the experimental measurements (see Table~\ref{tab:mom_FM}), but also results in the exchange interactions of the system preferring  a ferrimagnetic alignment. Even more the exchange interactions can be severely suppressed when the Hubbard $U$ is used. For example, for Co$_2$MnGe in ASA the dominant interaction is between the Co-Mn moments, in LSDA the obtained value is 0.79 mRy, while in LSDA$+U$[FLL] is reduced to 0.34 mRy, also, the nearest neighbour Co$_1$-Co$_2$ exchange interaction changes from ferromagnetic to antiferromagnetic when going from LSDA to LSDA$+U$[FLL] which lead the low values obtained for the spin wave stiffness. As will be discussed below also for the low $T_c$ for some of these systems.

It is important to notice, that the systems that exhibit the largest deviation from the experimental values, are usually those that under a certain exchange correlation potential and potential geometry loose their half-metallic character. Such effect are specially noticeable when one compares LSDA$+U$[FLL] results in ASA and FP, where half-metallicity is more easily lost in ASA due to the fact that the pseudogap is much smaller under this approximation than under FP (see Fig.~\ref{fig:DOS_CFS}). In general, it is important to notice that under ASA the geometry of the potential is imposed, that is non-spherical contributions to the potential are neglected. While this has been shown to be very successful to describe many properties, it does introduce an additional approximation which can lead to an ill treatment of the properties of some systems. Hence, care must be placed when one is considering an ASA treatment for the potential geometry, since it can lead to large variations of the exchange interactions and thus is one of the causes of the large spread on the values observed in Table~\ref{tab:jxc_HeuslerFM} for the exchange stiffness and in Table~\ref{tab:Tc_HeuslerFM} for the Curie temperature.

One of the key factors behind the small values of the spin stiffness for Co$_2$FeSi and Co$_2$FeGe, in comparison with the rest of the Co$_2$FeZ family, lies in  the fact that in LSDA and GGA an antiferromagnetic long-range Fe-Fe interaction is present (see Fig.~\ref{fig:Jij_CFS_app} in Appendix~\ref{app:Jij}). As the magnitude of the Fe-Fe interaction decreases the exchange stiffness increases, e.g. as in LSDA$+U$[AMF] with a full potential scheme. These exchange interactions are one of the factors behind the reduced value of the stiffness, this is evident when comparing with Co$_2$FeAl, which while having similar nearest neighbour Co-Fe exchange interactions, overall displays a much larger spin wave stiffness for most of the studied exchange correlation potentials.

Using LSDA+DMFT$\left[\Sigma\left(0\right)\right]$ for Co$_2$MnSi and Co$_2$FeSi, the obtained stiffness is 580 $\text{meV}\text{\AA}^2$ and 280 $\text{meV}\text{\AA}^2$ respectively, whilst in LSDA+DMFT[FLL] for Co$_2$MnSi the stiffness is 630  $\text{meV}\text{\AA}^2$ and for Co$_2$FeSi is 282 $\text{meV}\text{\AA}^2$. As can be seen for Co$_2$MnSi there is a good agreement between the KKR LSDA$+U$[FLL], the FP-LMTO LSDA+DMFT[FLL] and the experimental values.

The agreement with experiments is particularly good when correlation effects are considered as in the LSDA+DMFT$\left[\Sigma\left(0\right)\right]$ approach. On the other hand, for Co$_2$FeSi the spin wave stiffness is severely underestimated which is once again consistent with what is shown in Table~\ref{tab:jxc_HeuslerFM}.

\begin{table*}[]
\caption{Summary of the spin wave stiffness, $D$ for $\textrm{Co}_2\textrm{Mn}\textrm{Z}$ and $\textrm{Co}_2\textrm{Fe}\textrm{Z}$ with $\textrm{Z}=\left( \text{Al},\text{Si},\text{Ga},\text{Ge}\right)$. For the $\textrm{Co}_2\textrm{Mn}\textrm{Z}$ family both LSDA and GGA exchange correlation potentials yield values close to the experimental measurements. However, for the $\textrm{Co}_2\textrm{Fe}\textrm{Z}$ family a larger data spread is observed. The symbol $^*$ implies that the ground state for these systems was found to be Ferri-magnetic from Monte-Carlo techniques and the critical temperature presented here is calculated from the ferri-magnetic ground state.\label{tab:jxc_HeuslerFM}}
\begin{tabular}{l c c c c c c c c}
 \toprule
 \textbf{Quantity}										&$\text{Co}_2\text{Mn}\text{Al}$	& $\text{Co}_2\text{Mn}\text{Ga}$	& $ \text{Co}_2 \text{Mn}\text{Si}$	& $\text{Co}_2\text{Mn}\text{Ge}$ 	 &$\text{Co}_2\text{Fe}\text{Al}$	& $\text{Co}_2\text{Fe}\text{Ga}$	& $ \text{Co}_2 \text{Fe}\text{Si}$ 	& $\text{Co}_2\text{Fe}\text{Ge}$\\ \hline
 $D_\text{LDA}^{ASA}$ $[\text{meV}\text{\AA}^2]$			& 282							& 291								& 516								& 500			            		& 644							& 616							& 251								& 206							\\ 
 $D_\text{GGA}^{ASA}$ $[\text{meV}\text{\AA}^2]$			& 269							& 268								& 538								& 515							& 675							& 415							& 267								& 257							\\
 $D_{\text{LDA}+U}^{ASA}$ FLL $[\text{meV}\text{\AA}^2]$	& 29	$^*$							& 487$^*$							& 205								& 94 							& 289							& 289							& 314								& 173							\\ 
 $D_{\text{LDA}+U}^{ASA}$ AMF $[\text{meV}\text{\AA}^2]$	& 259							& 318								& 443				                & 417							& 553							& 588							& 235								& 214							\\ 
 $D_\text{LDA}^{FP}$ $[\text{meV}\text{\AA}^2]$			& 433							& 405								& 613   				                & 624							& 692							& 623							& 223								& 275							\\ 
 $D_\text{GGA}^{FP}$ $[\text{meV}\text{\AA}^2]$			& 483							& 452								& 691		    		                & 694	    						& 740							& 730							& 323								& 344							\\ 
 $D_{\text{LDA}+U}^{FP}$ FLL $[\text{meV}\text{\AA}^2]$	& 447							& 400								& 632				                & 577							& 652							& 611							& 461								& 436							\\ 
 $D_{\text{LDA}+U}^{FP}$ AMF $[\text{meV}\text{\AA}^2]$	& 216							& 348								& 583				                & 579							& 771							& 690							& 557								& 563							\\ 
 $D_\text{exp}$ $[\text{meV}\text{\AA}^2]$				& 190~\cite{JAP106113907}		& 264~\cite{6028176}					& 575~\cite{0022-3727-42-8-084005}-534~\cite{JAP111023912}&413~\cite{1.3296350}				& 370~\cite{00223727428084004}		& 496~\cite{002237273715001}			& 715~\cite{0022-3727-42-23-232001} 						&---            \\ \toprule
\end{tabular}

\end{table*}

Using the calculated exchange interactions, the critical temperature, $T_c$, for each system can be calculated. Using the ASA, the $T_c$ of both the $\text{Co}_2\text{Mn}\text{Z}$ and Co$_2$FeZ systems is consistently underestimated with respect to experimental results, as shown in Table~\ref{tab:Tc_HeuslerFM}. The same underestimation has been observed in previous theoretical studies~\cite{PhysRevB.89.094410}, for systems such as Co$_2$Fe(Al,Si) and Co$_2$Mn(Al,Si). However, using a full potential scheme instead leads to Curie temperatures in better agreement with the experimental values, specially when the exchange correlation potential is considered to be given by the GGA (see Table~\ref{tab:Tc_HeuslerFM}). Such observation is consistent with what was previously mentioned, regarding the effect of the ASA treatment on the spin wave stiffness and magnetic moments, where in certain cases, ASA was found to not be the best treatment to reproduce the experimental measurements. As mentioned above, this is strongly related to the fact that in general ASA yields a smaller pseudogap in the half-metallic materials, leading to modification of the exchange interactions. Thus, in general, a full potential approach seems to be able to better describe the magnetic properties in the present systems, since the pseudogap around the Fermi energy is better described in a FP approach for a given choice of exchange correlation potential. 

The inclusion of correlation effects for the Co$_2$FeZ family, lead to an increase of the Curie temperature, as for the spin stiffness. This is related to the enhancement of the interatomic exchange interactions as exemplified in the case of Co$_2$FeSi. However, the choice of DC once more is shown to greatly influence the magnetic properties. For the Co$_2$FeZ family, AMF results in much larger $T_c$ than the FLL scheme, whilst for Co$_2$MnZ the differences are smaller, with the exception of Z=Al. All these results showcase how important a proper description of the pseudogap region is in determining the magnetic properties of the system. 

Another observation, is the fact that even if a given combination of exchange correlation potential and geometrical treatment of the potential can yield a value of $T_c$ in agreement with experiments, it does not necessarily means that the spin wave stiffness is correctly predicted (see Table~\ref{tab:jxc_HeuslerFM} and Table~\ref{tab:Tc_HeuslerFM}).

\begin{table*}[]
\caption{Summary of the critical temperature for $\textrm{Co}_2\textrm{Mn}\textrm{Z}$ and $\textrm{Co}_2\textrm{Fe}\textrm{Z}$ with $\textrm{Z}=\left( \text{Al},\text{Si},\text{Ga},\text{Ge}\right)$, with different exchange correlation potentials and shape of the potentials. The symbol $^*$ implies that the ground state for these systems was found to be Ferri-magnetic from Monte-Carlo techniques and the critical temperature presented here is calculated from the ferri-magnetic ground state.\label{tab:Tc_HeuslerFM}}
\begin{tabular}{ l c c c c c c c c}
 \toprule
 \textbf{Quantity}                              &$\text{Co}_2\text{Mn}\text{Al}$	& $\text{Co}_2\text{Mn}\text{Ga}$	& $ \text{Co}_2 \text{Mn}\text{Si}$ 	& $\text{Co}_2\text{Mn}\text{Ge}$ 	 &$\text{Co}_2\text{Fe}\text{Al}$	& $\text{Co}_2\text{Fe}\text{Ga}$	& $ \text{Co}_2 \text{Fe}\text{Si}$ 	& $\text{Co}_2\text{Fe}\text{Ge}$\\ \hline
 $T_c^\text{LDA}$ ASA [K]	                    	& 360					        	& 350								& 750					                & 700	            				& 913				            	& 917			                	& 655								& 650								\\
 $T_c^\text{GGA}$ ASA [K]	                    	& 350					        	& 300					            & 763					                & 700    	      			& 975       			            	& 973				            	& 800								& 750								\\ 
 $T_c^{\text{LDA}+U}$ $\text{ASA}_\text{FLL}$ [K]& 50$^*$				            	& 625$^*$							& 125									& 225  	            				& 575							& 550							& 994								& 475								\\ 
 $T_c^{\text{LDA}+U}$ $\text{ASA}_\text{AMF}$ [K]& 325  				            	& 425			                		& 650				                		& 600   	            				& 950							& 950							& 650								& 625								\\ 
 $T_c^\text{LDA}$ FP [K]							& 525				            	& 475								& 875									& 825  	            				& 1050							& 975							& 750								& 750								\\ 
 $T_c^\text{GGA}$ FP [K]							& 600				            	& 525								& 1000									& 925           				&  1150      			        	& 1100							& 900								& 875								\\ 
 $T_c^{\text{LDA}+U}$ $\text{FP}_\text{FLL}$ [K]	& 525							& 475								& 950									& 875  	            				& 1050							& 975							& 1050								& 1075	            				   	\\ 
 $T_c^{\text{LDA}+U}$ $\text{FP}_\text{AMF}$ [K]	& 450				            	& 450								& 1000									& 875  	            				&1275							& 1225							& 1450								& 1350	            				   	\\ 
 $T_c^\text{exp}$ [K]							& 697~\cite{PhysRevB.89.094410}	& 694					            & 985~\cite{PhysRevB.74.104405}		    & 905            				& 1000~\cite{1.4863398}    		& 1093~\cite{10.1063/1.3536637} 	& 1100~\cite{PhysRevB.72.184434}		& 981~\cite{5257105}				\\ \toprule

\end{tabular}

\end{table*}

When considering the LSDA+DMFT[$\Sigma(0)$] scheme, critical temperatures of 688 K and 663 K are obtained for Co$_2$MnSi and Co$_2$FeSi, respectively. 
Thus, the values of the $T_c$ are underestimated in comparison with the LSDA+$U$ or LSDA results. 
The reason for such behaviour becomes clear when one looks directly on the $J_{ij}$'s, computed with the different schemes, which are shown in Appendix~\ref{app:Jij}.
These results suggest that taking into account the dynamical correlations (LSDA+DMFT[$\Sigma(0)$]) slightly suppresses most of the $J_{ij}$'s as compared to the LSDA outcome.
This is an expected result, since the employed choice of DC correction preserves the exchange splitting obtained in LSDA, while the dynamical self-energy, entering the Green's function, tends to lower its magnitude.
Since these two quantities are the key ingredients defining the strength of the exchange couplings, the $J_{ij}$'s obtained in DMFT are very similar to those of LSDA (see e.g. Refs.~\onlinecite{jij-yar,TM-surf-jijs}).
The situation is a bit different if one employs FLL DC, since an additional static correction enhances the local exchange splitting.\footnote{This manifests itself in the larger magnetic moment values, as was discussed above.}
For instance, in case of Co$_2$MnSi the LSDA+DMFT[FLL] scheme provided a $T_c$ of  764 K, which is closer to the experiment.
The consistently better agreement of the LSDA$+U$[FLL] and LSDA+DMFT[FLL] estimates of the $T_c$ with experimental values might indicate that explicit account for static local correlations is important for the all considered systems.

\begin{figure}
    \centering
    \includegraphics[width=\columnwidth]{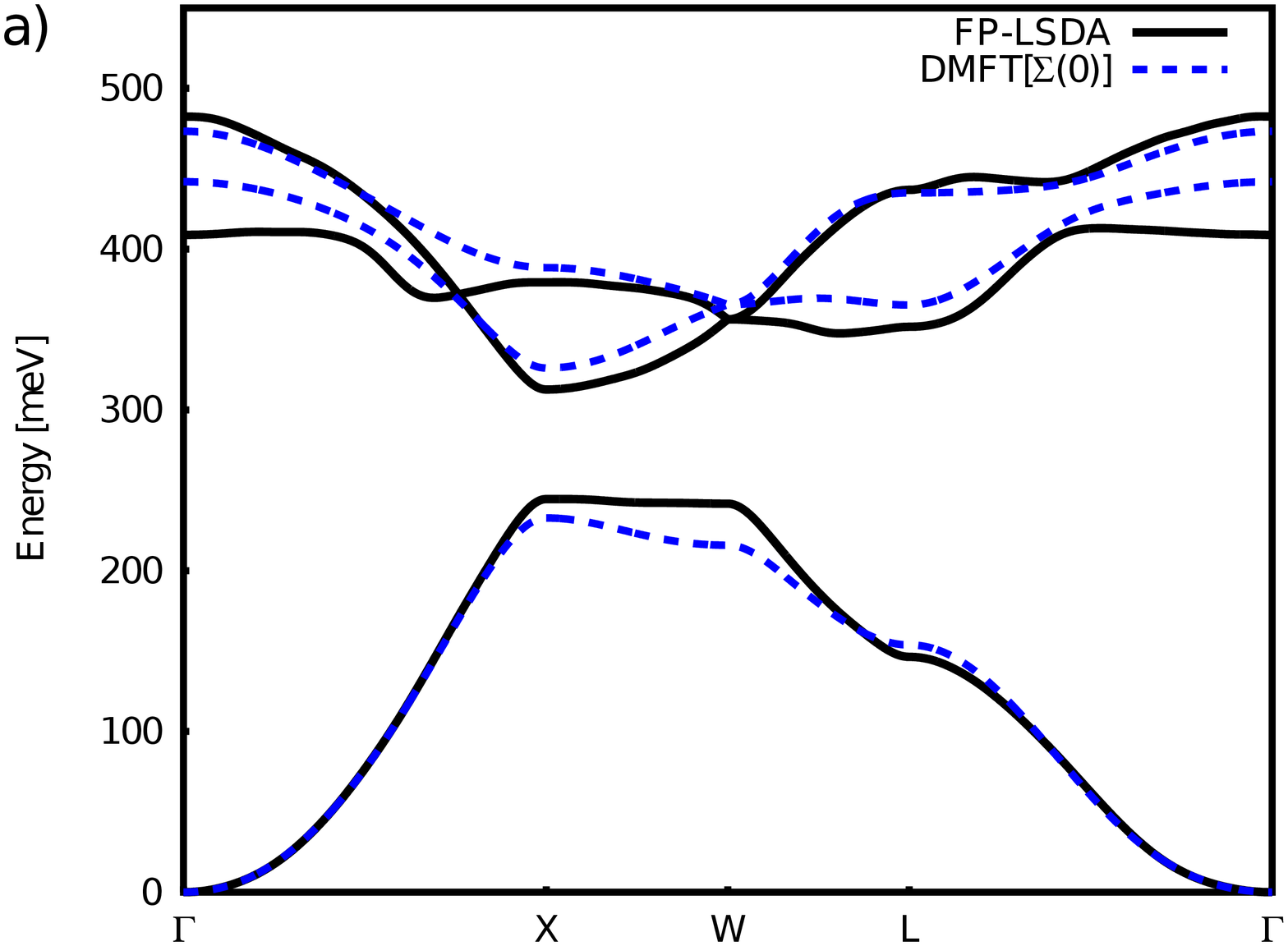}\\
    \includegraphics[width=\columnwidth]{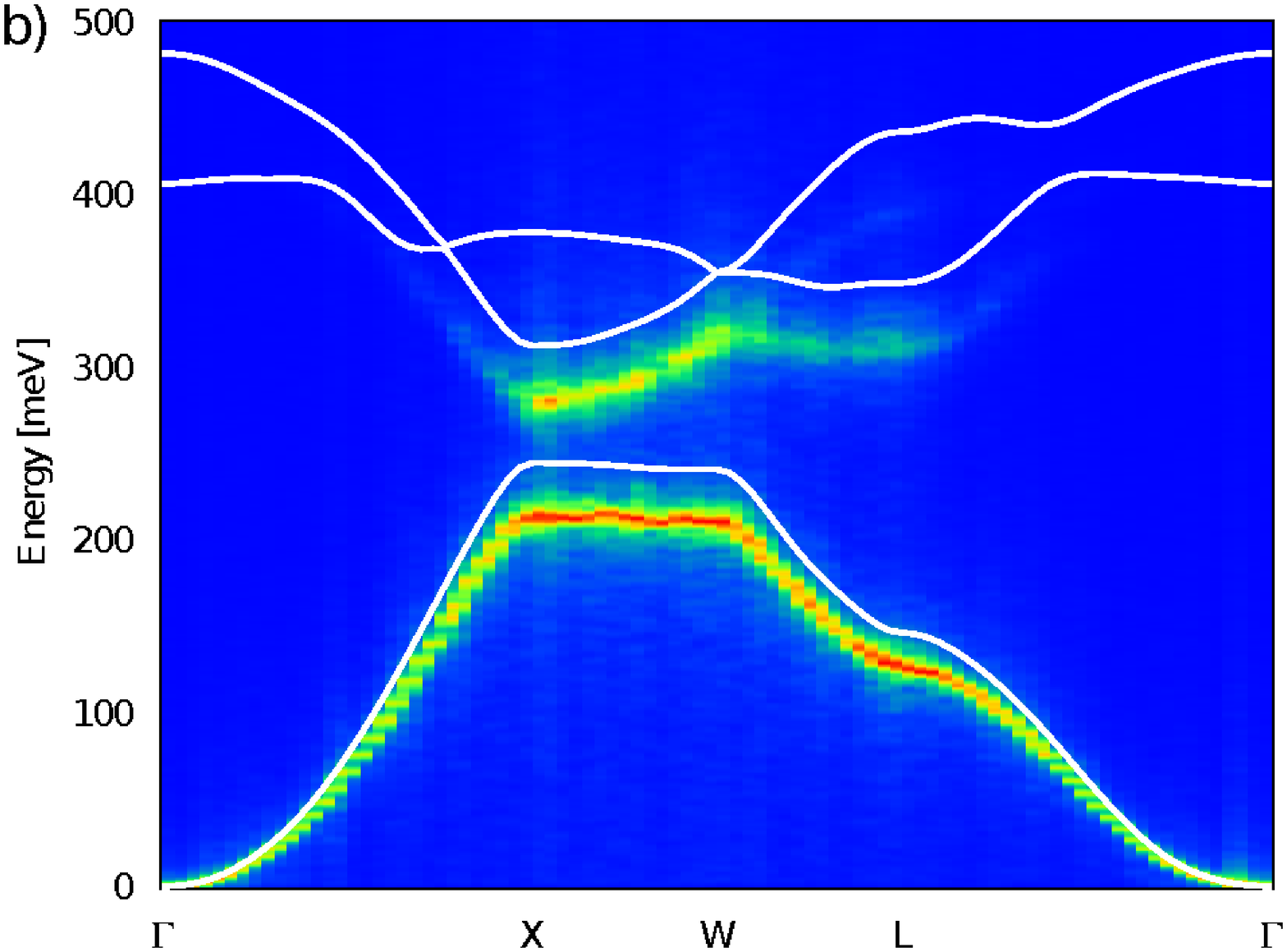}
    \caption{(Color online) a) Adiabatic magnon spectra for $\text{Co}_2\text{Fe}\text{Si}$ for different exchange correlation potentials. In the case of FP-LSDA and LSDA+DMFT[$\Sigma\left(0\right)$] the larger deviations are observed in the case of high energies, with the DMFT curve having a lower maximum than the LSDA results. In b) a comparison of the adiabatic magnon spectra (solid lines) with the dynamical structure factor $S\left(\mathbf{q},\omega\right)$ at $T= 300 \text{ K}$, when the shape of the potential is considered to be given by the atomic sphere approximation and the exchange correlation potential to be given by LSDA, some softening can be observed due to temperature effects specially observed at higher q-points.\label{fig:AMS}}
\end{figure}

Using the calculated exchange interactions, it is also possible to determine the adiabatic magnon spectra (AMS). In Fig.~\ref{fig:AMS} is shown the effect that different exchange correlation potentials have over the description of the magnon dispersion relation  of Co$_2$FeSi is shown. The most noticeable effect between different treatments of the exchange correlation potential is shifting the magnon spectra, while its overall shape seems to be conserved. This is a direct result from the enhancement of nearest neighbour interactions (see Fig.~\ref{fig:Jij_CFS_app}).

When comparing the AMS treatment with the dynamical structure factor, $S\left(\mathbf{q},\omega\right)$, at $T=300 \text{ K}$ and damping parameter $\alpha_\text{LSDA}=0.004$, obtained from first principles calculations (details explained in section~\ref{sec:DAMPING}), a good agreement at the long wavelength limit is found. However, a slight softening can be observed compared to the AMS. Such differences can be explained due to temperature effects included in the spin dynamics simulations. Due to the fact that the critical temperature of the system is much larger than $T=300 \text{ K}$ (see Table~\ref{tab:Tc_HeuslerFM}), temperature effects are quite small. The high energy optical branches are also softened and in general are much less visible. This is expected since the correlation was studied using only vectors in the first Brillouin zone and as has been shown in previous works~\cite{0953-8984-27-24-243202}, a phase shift is sometimes necessary to properly reproduce the optical branches, implying the need of vectors outside the first Brillouin zone. Also, Stoner excitations dealing with electron-hole excitations are not included in this approach, which result in the Landau damping which affects the intensity of the optical branches. Such effects are not captured by the present approach, but can be studied by other methods such as time dependent DFT~\cite{PhysRevLett.102.247206}. The shape of the dispersion relation along the path $\Gamma-X$ also corresponds quite well with previous theoretical calculations performed by K\"ubler~\cite{kubler2000theory}.

\subsection{Ferrimagnetic $\text{Mn}_2\text{V}\text{Z}$ with $\textrm{Z}=\left( \text{Al},\text{Si},\text{Ga},\text{Ge}\right)$}
As mentioned above, the Mn based $\text{Mn}_2\text{V}\text{Z}$ full Heusler family has a ferrimagnetic ground state, with the Mn atoms orienting parallel to each other and anti-parallel with respect to the V moments. For all the studied systems the Mn-Mn nearest neighbour exchange interactions dominates. In Table~\ref{tab:Tc_HeusV} the obtained spin wave stiffness, $D$, and critical temperature $T_c$ are shown. For $\text{Mn}_2\text{V}\text{Al}$, it can be seen that the spin wave stiffness is trend when compared to the experimental value. The same underestimation can be observed in the critical temperature. For Mn$_2$VAl, one may notice that the best agreement with  experiments is obtained for GGA in FP. An interesting aspect of the high $T_c$ observed in these materials is the fact that the magnetic order is stabilized due to the anti-ferromagnetic interaction between the Mn and V sublattices, since the Mn-Mn interaction is in general much smaller than the Co-Co, Co-Mn and Co-Fe interactions present in the previously studied ferromagnetic materials.

\begin{table}[]
\caption{Summary of the spin wave stiffness, $D$, and the critical temperature for $\textrm{Mn}_2\textrm{V}\textrm{Z}$ with $\textrm{Z}=\left( \text{Al},\text{Si},\text{Ga},\text{Ge}\right)$ for different shapes of the potential and exchange correlation potentials.\label{tab:Tc_HeusV}}
\begin{tabular}{l c c c c}
 \toprule
 \textbf{Quantity}								&$\text{Mn}_2\text{V}\text{Al}$	& $\text{Mn}_2\text{V}\text{Ga}$	& $ \text{Mn}_2 \text{V}\text{Si}$	& $\text{Mn}_2\text{V}\text{Ge}$ 	\\ \hline
 $D_\text{LDA}^{ASA}$ $[\text{meV}\text{\AA}^2]$	& 314		                    	& 114							& 					                	& 147		\\ 
 $D_\text{GGA}^{ASA}$ $[\text{meV}\text{\AA}^2]$	& 324		                    	& 73				                	& 					                	& 149		\\ 
 $D_\text{LDA}^{FP}$ $[\text{meV}\text{\AA}^2]$	& 421		                    	& 206							& 					                	& 191		\\ 
 $D_\text{GGA}^{FP}$ $[\text{meV}\text{\AA}^2]$	& 415		                    	& 91								& 					                	& 162		\\
 $D_\text{exp}$ $[\text{meV}\text{\AA}^2]$		& 534~\cite{Umetsu2015890}      	& ---							& ---                               	& ---        \\ 
 $T_c^\text{LDA}$ ASA [K]						& 275				            	& 350   		                    	& 150				                	& 147		\\
 $T_c^\text{GGA}$ ASA [K]						& 425				            	& 425 				            & 250				                	& 250		\\ 
 $T_c^\text{LDA}$ FP [K]							& 425   				            	& 450			                	& 200				                	& 200	    \\ 
 $T_c^\text{GGA}$ FP [K]							& 600				            	& 500				            	& 350				                	& 350	    \\ 
 $T_c^\text{exp}$ [K]							& 768~\cite{Umetsu2015890}     	& 783~\cite{RameshKumar20082737}	& ---	                            	& ---	    \\  \toprule
\end{tabular}
\end{table}

For these systems it can be seen that in general the FP description yields $T_c$'s which are in better agreement with experiment, albeit if the values are still underestimated. As for the Co based systems the full potential technique improves the description of the pseudogap, it is important to notice that for most systems both in ASA and FP the half-metallic character is preserved. However, the density of states at the Fermi level changes which could lead to changes in the exchange interactions.

\begin{figure}
    \centering
    \includegraphics[width=\columnwidth]{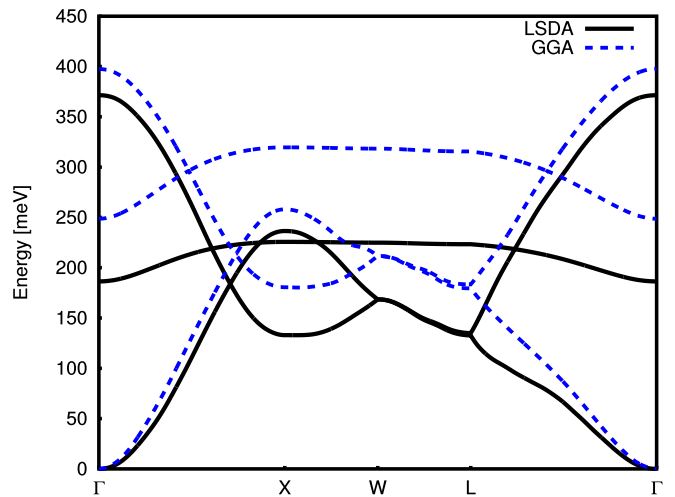}
    \caption{(Color online) Adiabatic magnon dispersion relation for $\text{Mn}_2\text{V}\text{Al}$ when different exchange correlation potentials are considered. In general only a shift in energy is observed when considering LSDA or GGA with the overall shape being conserved.\label{fig:AMS_Mn2VAl}}
\end{figure}

As for the ferromagnetic systems one can calculate the magnon dispersion relation and it is reported in Fig.~\ref{fig:AMS_Mn2VAl} for $\text{Mn}_2\text{V}\text{Al}$. A comparison with Fig.~\ref{fig:AMS} illustrates some of the differences between the dispersion relation of a ferromagnet and of a ferrimagnetic material. In Fig.~\ref{fig:AMS_Mn2VAl} some overlap between the acoustic and optical branches is observed, as well as a quite flat dispersion relation for one of the optical branches.
Such an effect is not observed in the studied ferromagnetic cases. In general the different exchange correlation potentials only tend to shift the energy of the magnetic excitations, while the overall shape of the dispersion does not change noticeably, which is consistent with what was seen in the ferromagnetic case.

The observed differences between the LSDA and GGA results in the small \textbf{q} limit, corresponds quite well with what is observed in Table~\ref{tab:Tc_HeusV}, where the spin wave stiffness for GGA with the potential given by ASA is somewhat larger than the LSDA case. This is directly related to the observation that the nearest neighbour Mn-Mn and Mn-V interactions are large in GGA than in LSDA. Again, such observation is tied to the DOS at the Fermi level, since Mn$_2$VAl is not half-metallic in LSDA, on the other hand in GGA the half-metallic state is obtained (see Table.~\ref{tab:mom_Ferri}.

\section{Gilbert damping\label{sec:DAMPING}}
\begin{figure}
\centering
\includegraphics[width=\columnwidth]{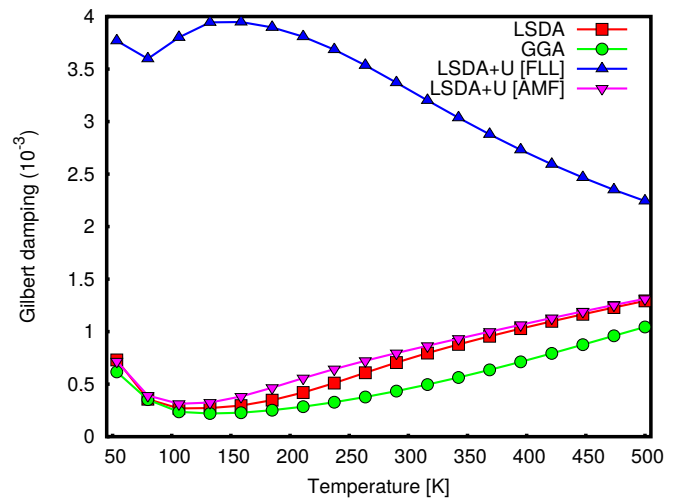}
\caption{(Color online) Temperature dependence of the Gilbert damping for $\text{Co}_2\text{Mn}\text{Si}$ for different exchange correlation potentials. For LSDA, GGA and LSDA$+U$[AMF] exchange correlation potentials the damping increases with temperature, whilst for LSDA$+U$[FLL] the damping decreases as a function of temperature.\label{fig:CMS_DAMP_T}}
\end{figure}

The Gilbert damping is calculated for all the previously studied systems using ASA and a fully relativistic treatment. In Fig.~\ref{fig:CMS_DAMP_T}, the temperature dependence of the Gilbert damping for $\text{Co}_2\text{Mn}\text{Si}$ is reported for different exchange-correlation potentials. When correlation effects are neglected or included via the LSDA$+U$[AMF], the damping increases with temperature. On the other hand, in the LSDA$+U$[FLL] scheme, the damping decreases as a function of temperature, and its overall magnitude is much larger. Such observation can be explained from the fact that in this approximation a small amount of states exists at the Fermi energy in the pseudogap region, hence resulting in a larger damping than in the half-metallic cases(see Fig.~\ref{fig:DOS_CFS}c).

In general the magnitude of the damping, $\alpha_\text{LSDA}=7.4\times 10^{-4}$, is underestimated with respect to older experimental measurements at room temperature, which yielded values of $\alpha=\left[0.003-0.006\right]$~\cite{1.3456378} and $\alpha\sim 0.025$ for polycrystalline samples~\cite{Yilgin20072322}, whilst it agrees with previously performed theoretical calculations~\cite{0022-3727-48-16-164011}. Such discrepancy between the experimental and theoretical results could stem from the fact that in the theoretical calculations only the intrinsic damping is calculated, while in experimental measurements in addition extrinsic effects such as eddy currents and magnon-magnon scattering can affect the obtained values. It is also known that sample capping or sample termination, can have profound effects over the half-metallicity of Co$_2$MnSi~\cite{PhysRevLett.94.096402}. Recent experiments showed that ultra-low damping, $\alpha=7\times 10^{-4}$, for Co$_{1.9}$Mn$_{1.1}$Si can be measured when the capping is chosen such that the half-metallicity is preserved~\cite{PhysRevB.93.094417}, which is in very good agreement with the present theoretical calculations. 

\begin{figure}
\centering
\includegraphics[width=\columnwidth]{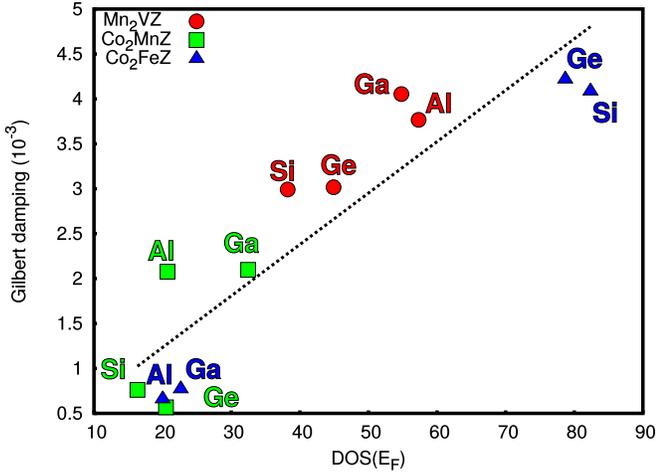}
\caption{(Color online) Gilbert damping for different Heusler alloys at $T=300\text{ K}$ as a function of density of states at the Fermi energy for LSDA exchange correlation potential. In general the damping increases as the density of states at the Fermi Energy increases (the dotted line is to guide the eyes).\label{fig:DAMP300_Heus}}
\end{figure}

In Fig.~\ref{fig:DAMP300_Heus}, the Gilbert damping at $T=300 \text{ K}$ for the different Heusler alloys as a function of the density of states at the Fermi level is presented. As expected, the increased density of states at the Fermi energy results in an increased damping. Also it can be seen that in general, alloys belonging to a given family have quite similar damping parameter, except for Co$_2$FeSi and  Co$_2$FeGe. Their anomalous  behaviour, stems from the fact that in the LSDA approach both Co$_2$FeSi and Co$_2$FeGe are not half-metals. Such clear dependence on the density of states is expected, since the spin orbit coupling is small for these materials, meaning that the dominating contribution to the damping comes from the details of the density of states around the Fermi energy~\cite{PhysRevB.87.014430,PhysRevB.91.104420}.

\subsubsection{Effects of substitutional disorder}
In order to investigate the possibility to influence the damping, we performed calculations for the chemically disordered Heusler alloys $\text{Co}_2\text{Mn}_{1-x}\text{Fe}_x \text{Si}$, $\text{Co}_2\text{Me}\text{Al}_{1-x} \text{Si}_x$ and $\text{Co}_2\text{Me}\text{Ga}_{1-x} \text{Ge}_x$ where $\text{Me}=\left(\text{Mn},\text{ Fe}\right)$. 

\begin{figure}
\centering
\includegraphics[width=\columnwidth]{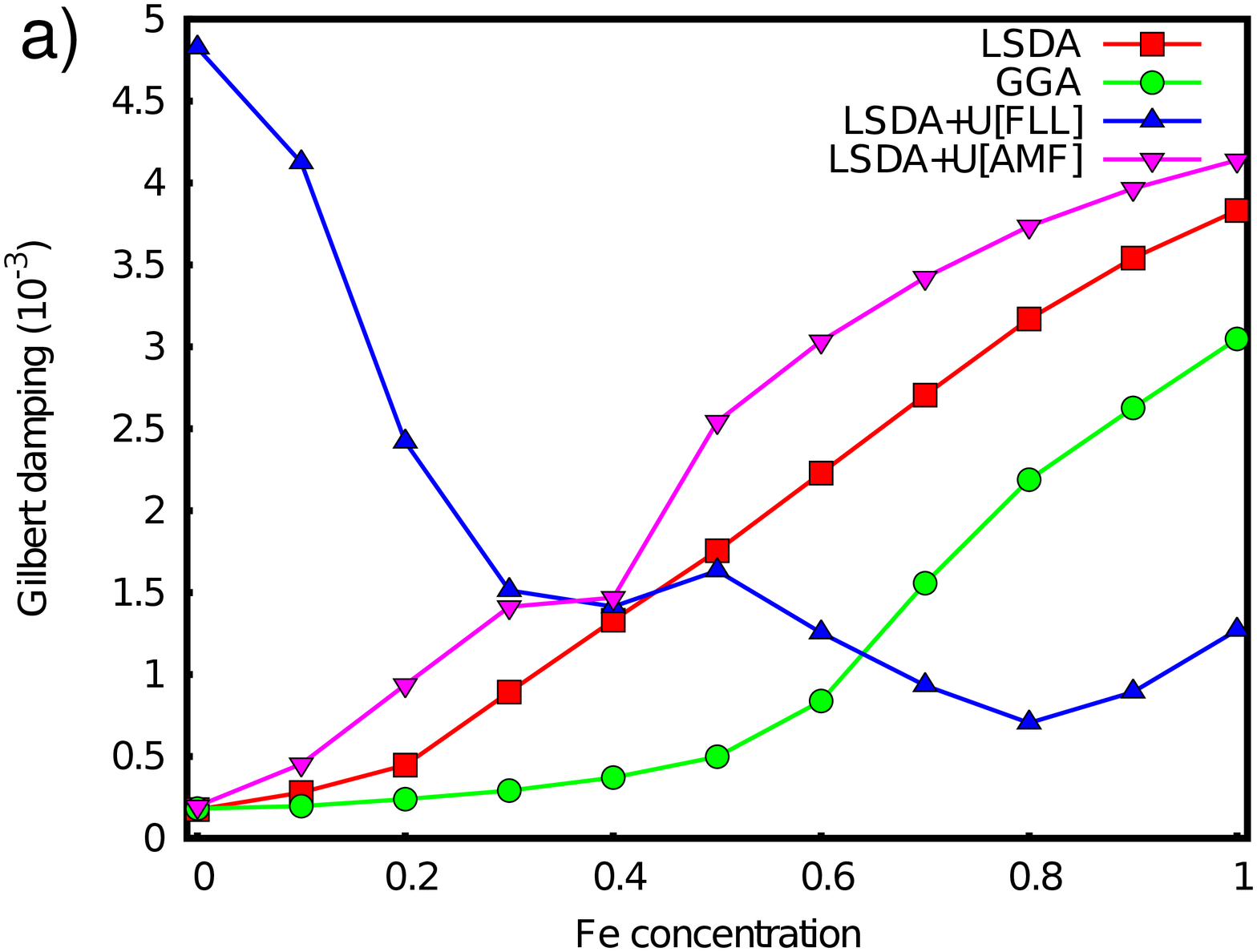}\\
\includegraphics[width=\columnwidth]{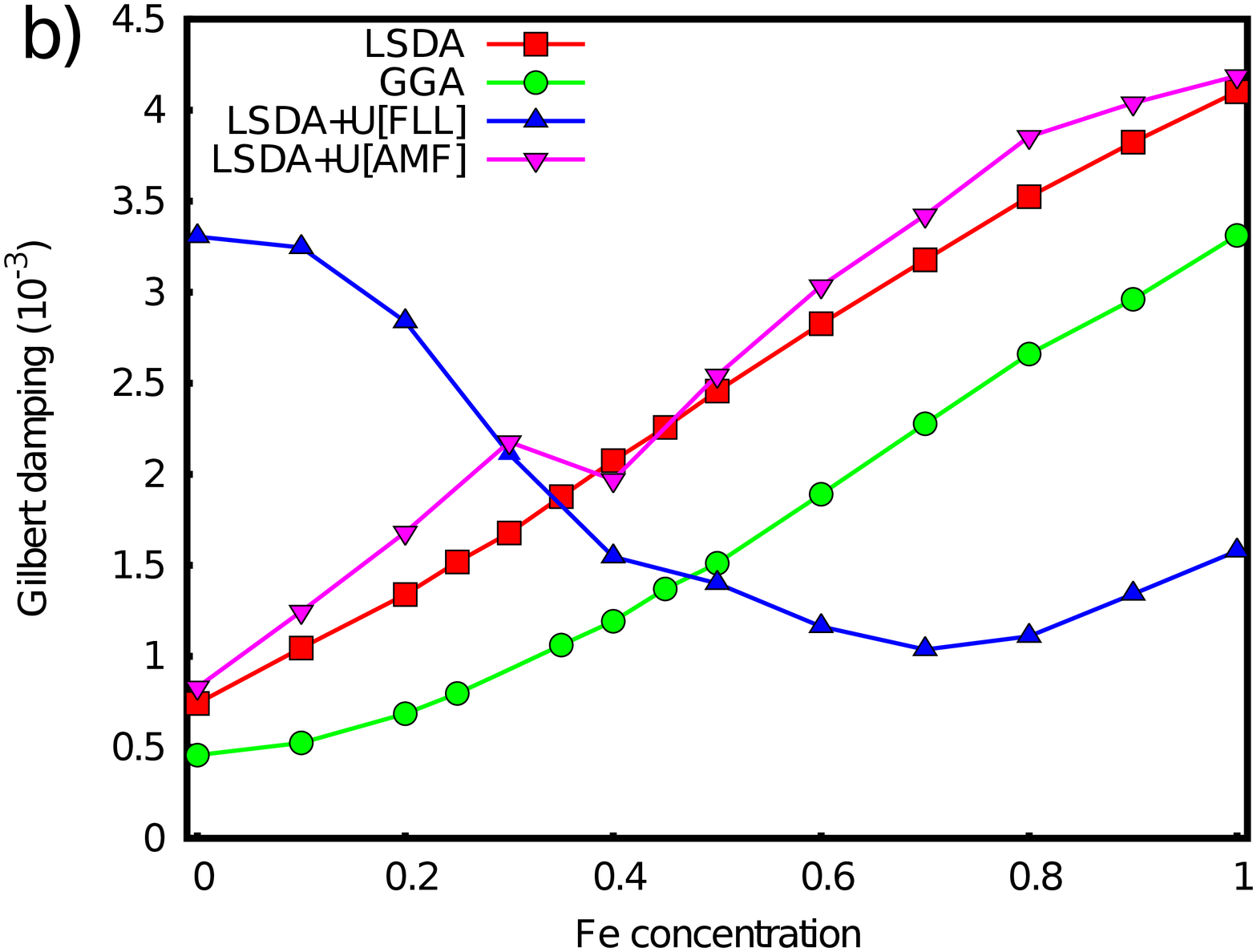}
\caption{(Color online) Gilbert damping for the random alloy $\text{Co}_2\text{Mn}_{1-x}\text{Fe}_x \text{Si}$ as a function of the Fe concentration at $T=300 \text{ K}$ when a) only atomic deisplacements are considered and b) when both atomic displacements and spin fluctuations are considered.\label{fig:Heusdampalloy}}
\end{figure}

Due to the small difference between the lattice parameters of Co$_2$MnSi and Co$_2$FeSi, the lattice constant is unchanged when varying the concentration of Fe. This is expected to play a minor role on the following results.
When one considers only atomic displacement contributions to the damping (see Fig.~\ref{fig:Heusdampalloy}a), the obtained values are clearly underestimated in comparison with the experimental measurements at room temperature\cite{APL94122504}. Under the LSDA, GGA and LSDA$+U$[AMF] treatments, the damping is shown to increase with increasing concentration of Fe. On the other hand, in LSDA$+U$[FLL] the damping at low concentrations of Fe is much larger than in the other cases, and it decreases with Fe concentration, until a minima is found at Fe concentration of $x \sim 0.8$. This increase can be related to the DOS at the Fermi energy, which is reported in Fig.~\ref{fig:DOS_CFS}c for Co$_2$MnSi. One can observe a small amount of states at $E_F$, which could lead to increased values of the damping in comparison with the ones obtained in traditional LSDA. As for the pure alloys, a general trend relating the variation of the DOS at the Fermi level and the damping with respect to the variation of Fe concentration can be obtained, analogous to the results shown in Fig.~\ref{fig:DAMP300_Heus}. 

When spin fluctuations are considered in addition to the atomic displacements contribution, the magnitude of the damping increases considerably, as shown in Fig.~\ref{fig:Heusdampalloy}b. This is specially noticeable at low concentrations of Fe. Mn rich alloys have a $T_c$ lower than the Fe rich ones, thus resulting in larger spin fluctuations at $T=300 \text{ K}$. The overall trend for LSDA and GGA is modified at low concentrations of Fe when spin fluctuations are considered, whilst for LSDA$+U$[FLL] the changes in the trends occur mostly at concentrations between $x=\left[0.3-0.8\right]$. An important aspect is the overall good agreement of LSDA, GGA and LSDA$+U$[AMF]. Instead results obtained in LSDA$+U$[FLL] stand out as different from the rest. This is is expected since as was previously mentioned the FLL DC  is not the most appropriate scheme to treat these systems. An example of such inadequacy can clearly be seen in Fig.~\ref{fig:Heusdampalloy}b for Mn rich concentrations, where the damping is much larger with respect to the other curves. As mentioned above, this could result from the appearance of states at the Fermi level. 

Overall the magnitude of the intrinsic damping presented here is smaller than the values reported in experiments~\cite{APL94122504}, which report values for the damping of Co$_2$MnSi of $\alpha\sim 0.005$ and $\alpha\sim 0.020$ for Co$_2$FeSi, in comparison with the calculated values of $\alpha_\text{LSDA}=7.4\times 10^{-4}$ and $\alpha_{LSDA}=4.1\times 10^{-3}$ for Co$_2$MnSi and Co$_2$FeSi, respectively. In experiments also a minimum at the concentration of Fe of $x\sim 0.4$ is present, while such minima is not seen in the present calculations. However, similar trends as those reported here (for LSDA and GGA) are seen in the work by Oogane and Mizukami~\cite{Oogane3037}. A possible reason behind the discrepancy between theory and experiment, could stem from the fact that as the Fe concentration increases, correlation effects also increase in relative importance. Such a situation cannot be easily described through the computational techniques used in this work, and will affect the details of the DOS at the Fermi energy, which in turn could modify the damping. Another important factor influencing the agreement between theory and experiments arise form the difficulties in separating extrinsic and intrinsic damping in experiments~\cite{PhysRevLett.90.227601}. This, combined with the large spread in the values reported in various experimental studies~\cite{1347-4065-46-3L-L205,7156903,Yilgin20072322}, points towards the need of improving both theoretical and experimental approaches, if one intends to determine the minimum damping attainable for these alloys with sufficient accuracy.

Up until now in the present work, disorder effects have been considered at the Y site of the Heusler structure. In the following chemical disorder will be considered on the Z site instead. Hence, the chemical structure changes to the type Co$_2$MeZ$^A_{1-x}$Z$^B_{x}$ (Me=Fe,Mn). The alloys Co$_2$MeAl$_x$Si$_{1-x}$ and Co$_2$MeGa$_x$Ge$_{1-x}$ are considered. The lattice constant for the off stoichiometric compositions is treated using Vegard's law~\cite{VegardLaw}, interpolating between the values given in Table~\ref{tab:mom_FM}.

\begin{figure}[h]
\centering
\includegraphics[width=\columnwidth]{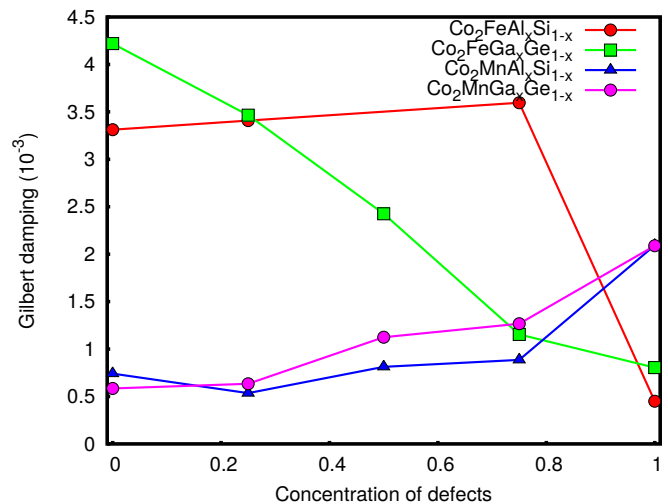}
\caption{(color online) Dependence of the Gilbert damping for the alloys Co$_2$MeAl$_x$Si$_{1-x}$ and Co$_2$MeGa$_x$Ge$_{1-x}$ with Me denoting Mn or Fe under the LSDA exchange correlation potential.\label{fig:CMALL}}
\end{figure}

In Fig.~\ref{fig:CMALL} the dependence of the damping on the concentration of defects is reported, as obtained in LSDA. For Co$_2$FeGa$_x$Ge$_{1-x}$ as the concentration of defects increases the damping decreases. Such a behaviour can be understood by inspecting the density of states at the Fermi level which follows the same trend, it is important to notice that Co$_2$FeGa is a half-metallic system, while Co$_2$FeGe is not (see table~\ref{tab:mom_FM}). On the other hand, for Co$_2$FeAl$_x$Si$_{1-x}$, the damping increases slightly with Al concentration, however, for the stoichiometric Co$_2$FeAl is reached the damping decreases suddenly, as in the previous case. This is a direct consequence of the fact that Co$_2$FeAl is a half metal and Co$_2$FeSi is not, hence when the half-metallic state is reached a sudden decrease of the damping is observed. For the Mn based systems, as the concentration of defects increases the damping increases, this stark difference with the Fe based systems. For Co$_2$MnAl$_x$Si$_{1-x}$ this is related to the fact that both Co$_2$MnAl and Co$_2$MnSi are half-metals in LSDA, hence, the increase is only related to the fact that the damping for Co$_2$MnAl is larger than the one of Co$_2$MnSi, it is also relevant to mention, that the trend obtained here corresponds quite well with what is observed in both experimental and theoretical results in Ref.~\cite{1.3456378}. A similar explanation can be used for the Co$_2$MnGa$_x$Ge$_{1-x}$ alloys, as both are half-metallic in LSDA. As expected, the half metallic Heuslers have a lower Gilbert damping than the other ones, as shown in Fig.~\ref{fig:CMALL}.

\section{Conclusions}
The treatment of several families of half-metallic Heusler alloys has been systematically investigated using several approximations for the exchange correlation potential, as well as for the shape of the potential. Special care has been paid to the calculation of their magnetic properties, such as the Heisenberg exchange interactions and the Gilbert damping. Profound differences have been found in the description of the systems depending on the choice of exchange correlation potentials, specially for systems in which correlation effects might be necessary to properly describe the presumed half-metallic nature of the studied alloy.

In general, no single combination of exchange correlation potential and potential geometry was found to be able to reproduce all the experimentally measured magnetic properties of a given system simultaneously. Two of the key contributing factors are the exchange correlation potential and the double counting scheme used to treat correlation effects. The destruction of the half-metallicity of any alloy within the study has profound effects on the critical temperature and spin wave stiffness. A clear indication of this fact is that even if the FLL double counting scheme may result in a correct description of the magnetic moments of the system, the exchange interactions may be severely suppressed. For the systems studied with DMFT techniques either minor improvement or results similar to the ones obtained from LSDA is observed. This is consistent with the inclusion of local $d-d$ screening, which effectively diminishes the strength of the effective Coulomb interaction with respect to LSDA$+U$ (for the same Hubbard parameter $U$). In general, as expected, the more sophisticated treatment for the geometrical shape of the potential, that is a full potential scheme, yields results closer to experiments, which in these systems, is intrinsically related to the description of the pseudogap region.

Finally, the Gilbert damping is underestimated with respect to experimental measurements, but in good agreement with previous theoretical calculations. One of the possible reasons being the difficulty from the experimental point of view of separating intrinsic and extrinsic contributions to the damping, as well as the strong dependence of the damping on the crystalline structure. A clear correlation between the density of states at the Fermi level and the damping is also observed, which is related to the presence of a small spin orbit coupling in these systems. This highlights the importance that half-metallic materials, and their alloys, have in possible spintronic and magnonic applications due to their low intrinsic damping, and tunable magnetodynamic variables. These results could spark interest from the experimental community due to the possibility of obtaining ultra-low damping in half-metallic Heusler alloys.

\section{Acknowledgements}
The authors acknowledge valuable discussions with M.I. Katsnelsson and A.I. Lichtenstein. The work was financed through the VR (Swedish Research Council) and GGS (G\"oran Gustafssons Foundation). O.E. acknowledges support form the KAW foundation (grants 2013.0020 and 2012.0031).  O.E. and A.B acknowledge eSSENCE. L.B acknowledge support from the Swedish e-Science Research Centre (SeRC). The computer simulations were performed on resources provided by the Swedish National Infrastructure for Computing (SNIC) at the National Supercomputer Centre (NSC) and High Performance Computing Center North (HPC2N).

\appendix
\counterwithin{figure}{section}
\counterwithin{table}{section}

\section{DOS from LSDA+DMFT\label{app:DOSDMFT}}
\begin{figure}[!t]
\centering
\includegraphics[width=0.80\columnwidth]{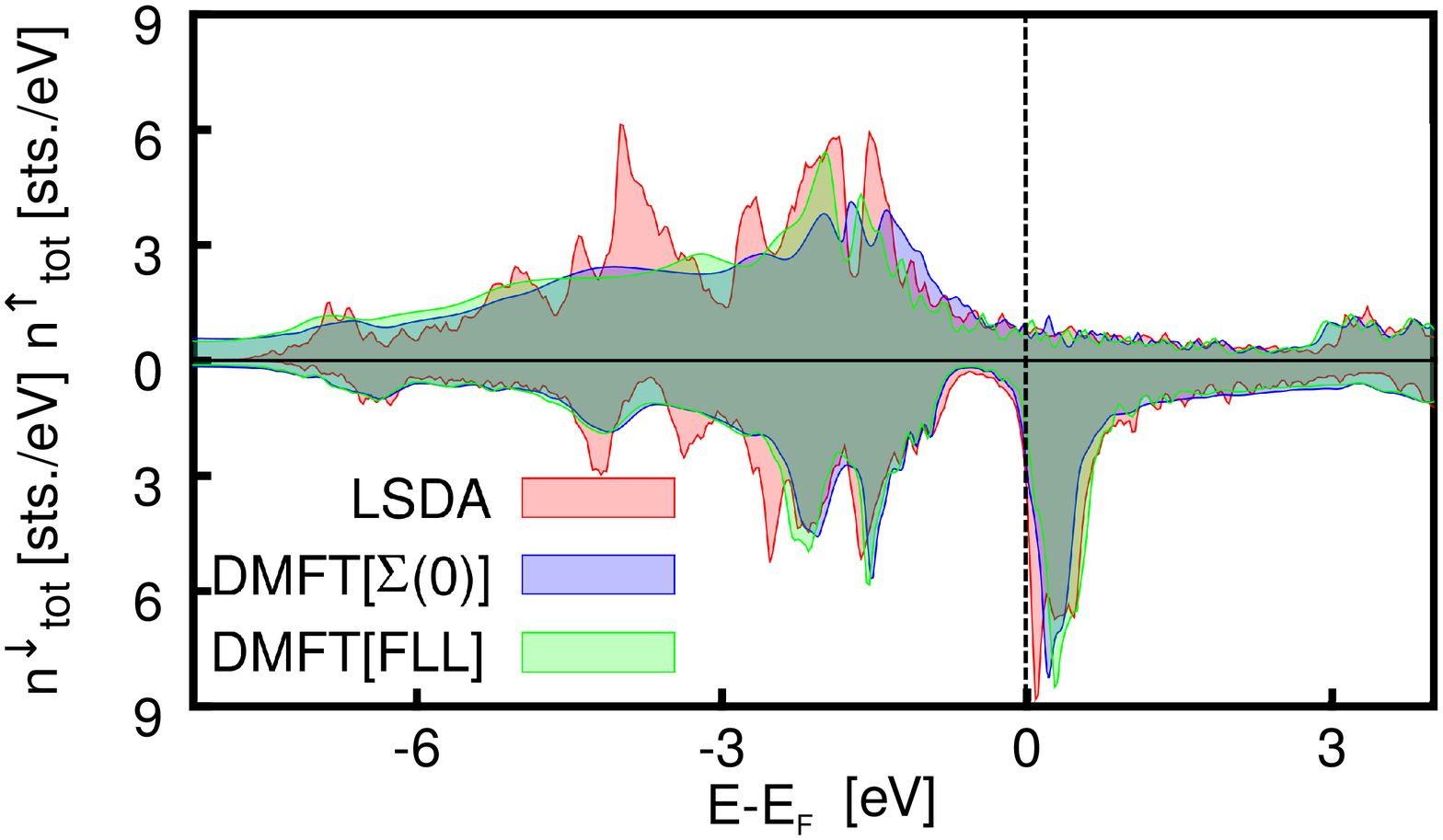}\\
\includegraphics[width=0.80\columnwidth]{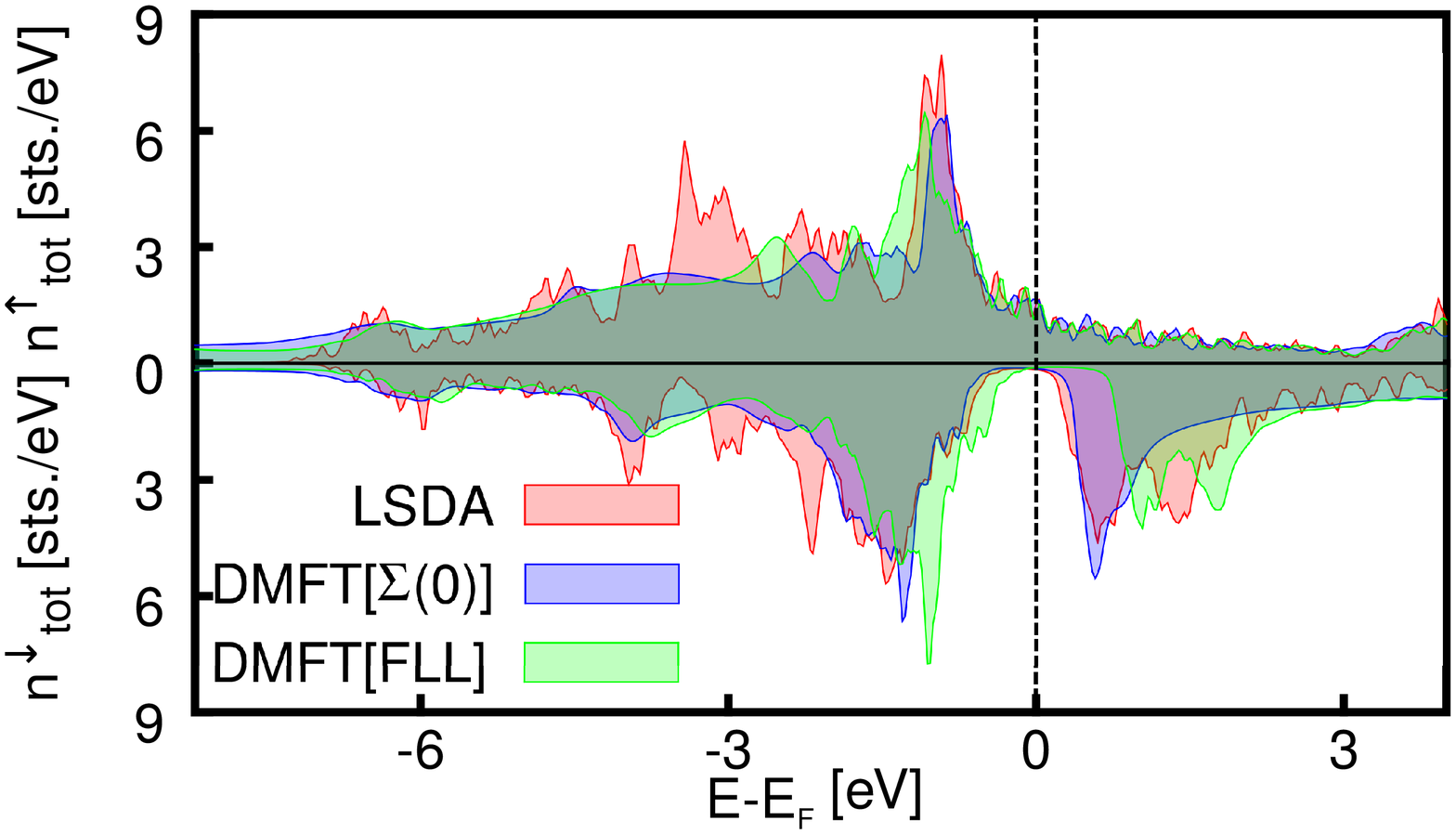}\\
\caption{(color online) DOS in Co$_2$FeSi (top panel) and Co$_2$MnSi (bottom panel) obtained in different computational setups.\label{fig:DOS_APP}}
\end{figure}

Here we show the DOS in Co$_2$MnSi and Co$_2$FeSi obtained from LSDA and LSDA+DMFT calculations. The results shown in Fig.~\ref{fig:DOS_APP} indicate that the DMFT increases the spin-down (pseudo-)gap in both Co$_2$FeSi and Co$_2$MnSi. In the latter case the shift of the bands is more pronounced. In Co$_2$FeSi it manifests itself in an enhanced value of the total magnetization. For both studied systems, the FLL DC results in relatively larger values of the gaps as compared with the \textquotedblleft$\Sigma(0)$\textquotedblright estimates. However, for the same choice of the DC this gap appears to be smaller in LSDA+DMFT than in LSDA+$U$. Present conclusion is valid for both Co$_2$FeSi and Co$_2$MnSi (see Fig.~\ref{fig:DOS_CFS} for comparison.)

\section{NQS in Co$_2$MnSi\label{app_NQS}}

\begin{figure}[!t]
\centering
\includegraphics[width=0.75\columnwidth]{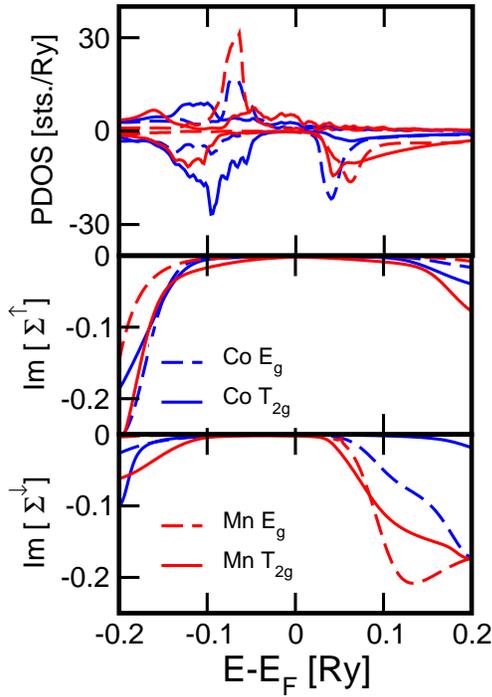}
\caption{(color online) Top panel: DOS in Co$_2$MnSi projected onto Mn and Co $3d$ states of different symmetry. Middle and bottom panels: Orbital-resolved spin-up and spin-down imaginary parts of the self-energy. The results are shown for the \textquotedblleft$\Sigma(0)$\textquotedblright DC.\label{fig:NQS_app}}
\end{figure}

Here we show the calculated spectral functions in Co$_2$MnSi obtained with LSDA+DMFT[$\Sigma(0)$] approach. As discussed in the main text, the overall shape of DOS is reminiscent of that obtained in LSDA.  However, a certain amount of the spectral weight appears above the minority-spin gap.
An inspection of the imaginary part of the self-energy in minority-spin channel, shown in the bottom panel of Fig.~\ref{fig:NQS_app}, suggests a strong increase of Mn spin-down contribution at the corresponding energies, thus confirming the non-quasiparticle nature of the obtained states. 
We note that the use of FLL DC formulation results in an enhanced spin-down gap which pushes the NQS to appear at even higher energies above $E_F$ (see Appendix~\ref{app:DOSDMFT}).

\section{Impact of correlation effects on the $J_{ij}$'s in Co$_2$MnSi and Co$_2$FeSi\label{app:Jij}}
\begin{figure}[!t]
\centering
\includegraphics[width=0.95\columnwidth]{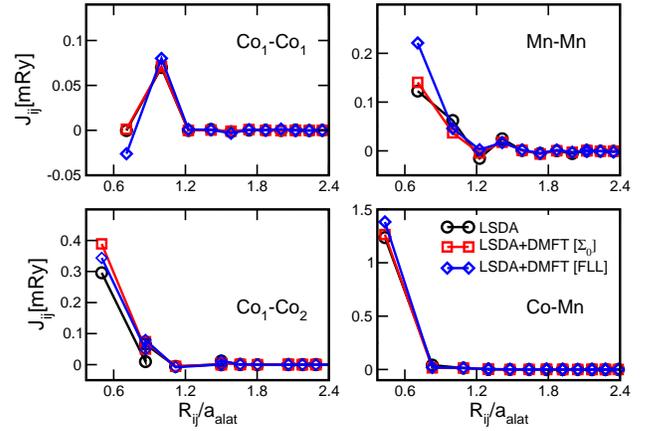}
\caption{(color online) The calculated exchange parameters in Co$_2$MnSi within LSDA and LSDA+DMFT for different choice of DC.\label{fig:Jij_app}}
\end{figure}

In this section we present a comparison of the exchange parameters calculated in the framework of the LSDA+DMFT using different DC terms. 
The calculated $J_{ij}$'s between different magnetic atoms within the first few coordination spheres are shown in Fig.~\ref{fig:Jij_app}.
One can see that the leading interactions which stabilize the ferromagnetism in these systems are the nearest-neighbour intra-sublattice couplings between Co and Fe(Mn) atoms and, to a lower extend, the interaction between two Co atoms belonging to the different sublattices.
This qualitative behaviour is obtained independently of the employed method for treating correlation effects and is in good agreement with prior DFT studies.
As explained in the main text, the LSDA and LSDA+DMFT[$\Sigma(0)$] results are more similar to each other, whereas most of the $J_{ij}$'s extracted from LSDA+DMFT[FLL] are relatively enhanced due to inclusion of an additional static contribution to the exchange splitting. This is also reflected in both values of the spin stiffness and the $T_c$.

\begin{table}
\centering
\caption{Orbital-resolved $J_{ij}$'s between the nearest neighbours in Co$_2$MnSi in mRy. In the case of Co$_1$-Co$_1$, the second nearest neighbour value is given, due to smallness of the first one. The results were obtained with LSDA.}
\label{orbital-resolved}
\begin{tabular}{ l c c c c c}
\toprule
				&  Total		&$E_g-E_g$	& $T_{2g}-T_{2g}$	& $E_{g}-T_{2g}$	& $T_{2g}-E_g$	\\
\hline
Co$_1$-Co$_1$	&    0.070	&  0.077  	&  -0.003   			& -0.002			&   -0.002		\\     
Co$_1$-Co$_2$	&    0.295	&  0.357		&  -0.058			& -0.002			&   -0.002		\\  
Co-Mn			&    1.237	&  0.422		&  -0.079			&  0.700			&    0.194		\\
Mn-Mn			&    0.124	& -0.082		&   0.118			&  0.044			&    0.044		\\
\toprule
\end{tabular}
\end{table}

In order to have a further insight into the details of the magnetic interactions in the system, we report here the orbital-resolved $J_{ij}$'s between the nearest-neighbours obtained with LSDA. 
The results, shown in Table.~\ref{orbital-resolved}, reveal few interesting observations.
First of all, all the $T_{2g}$-derived contributions are negligible for all the interactions involving Co atoms.
This has to do with the fact that these orbitals are practically filled and therefore can not participate in the exchange interactions.
As to the most dominant Co-Mn interaction, the $E_g-E_g$ and $E_{g}-T_{2g}$ contributions are both strong and contribute to the total ferromagnetic coupling. This is related to strong spin polarisation of the Mn-$E_g$ states.

\begin{figure}
\centering
\includegraphics[width=0.95\columnwidth]{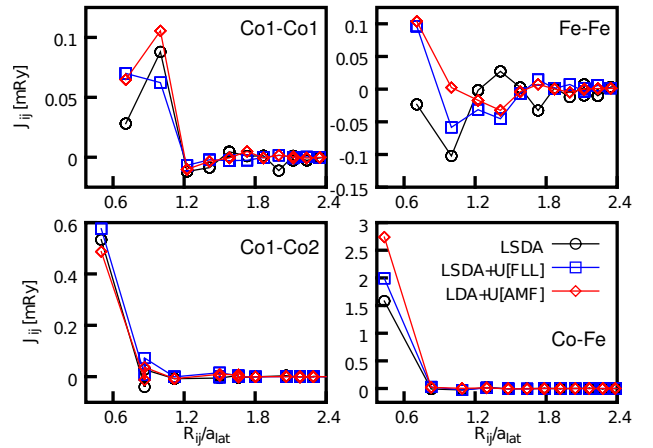}
\caption{\label{fig:Jij_CFS_app} Exchange interactions for Co$_2$FeSi within LSDA and LSDA$+U$ schemes and a full potential approach for different DC choices.}
\end{figure}

Correlation effects also have profound effects on the exchange interactions of Co$_2$FeSi. In particular, the Fe-Fe interactions can be dramatically changed when considering static correlation effects. It is specially noticeable how the anti-ferromagnetic exchange interactions can decrease significantly which can affect the exchange stiffness and the critical temperature as described in the main text. Also the long-range nature of the Fe-Fe interactions is on display, indicating that to be able to predict macroscopic variables from the present approach, attention must be paid to the cut-off range. As in Co$_2$MnSi, correlation effect do not greatly affect the Co-Co exchange interactions.

\FloatBarrier
\bibliographystyle{apsrev}

\end{document}